\documentclass[12pt, oneside]{article}   	
\usepackage[a4paper,top=2.5cm,bottom=2.5cm,left=2.5cm,right=2.5cm, heightrounded]{geometry}
\usepackage{graphicx}				
\usepackage{amssymb}

\usepackage{newtxtext}
\usepackage{newtxmath}
\usepackage{natbib}
\usepackage{hyperref}
\usepackage{epstopdf, epsfig}
\usepackage[utf8]{inputenc}
\usepackage{amsmath}
\usepackage{color}
\usepackage{listings}
\usepackage{float}
\usepackage{xcolor}
\hypersetup{
    colorlinks = false,
    urlcolor   = blue,
    citecolor  = black,
}

\newcommand{\RomanNumeralCaps}[1]
\linenumbers

\usepackage{graphicx}
\usepackage{appendix}
\usepackage{hyperref}
\usepackage{xcolor}
\usepackage{multirow}
\usepackage{float}
\usepackage{amsmath}
\usepackage{booktabs} 

\providecommand\bnabla{\boldsymbol{\nabla}}
\newcommand\Rey{\mbox{\textit{Re}}}  
\providecommand\bcdot{\boldsymbol{\cdot}}
\DeclareMathAlphabet\mathsfbi            {OT1}{cmss}{m}{sl}

\title{\textbf{Modelling spatiotemporal turbulent dynamics with the convolutional autoencoder echo state network}}

\author{Alberto Racca$^1$, Nguyen Anh Khoa Doan$^2$, and Luca Magri$^{3,1,4}$ \footnote{l.magri@imperial.ac.uk} \\
\small{$^1$ Department of Engineering, University of Cambridge,
Cambridge CB2 1PZ, UK} \\
\small{$^2$ Delft University of Technology, Faculty of Aerospace Engineering, 2629 HS, Delft, The Netherlands} \\
\small{$^3$ Imperial College London, Aeronautics Department,
 London, SW7 2AZ, UK} \\
\small{$^4$ The Alan Turing Institute,  NW1 2DB, London, UK}
}

\date{}

\begin{document}
\maketitle
\begin{abstract}
\noindent The spatiotemporal dynamics of turbulent flows is chaotic and difficult to predict. 
This makes the design of accurate and stable reduced-order models challenging.
The overarching objective of this paper is to propose a nonlinear decomposition of the turbulent state for a reduced-order representation of the dynamics. 
We divide the turbulent flow into a spatial problem and a temporal problem. 
First, we compute the latent space, which is the manifold onto which the turbulent dynamics live (i.e., it is a numerical approximation of the turbulent attractor). The latent space is found by a series of nonlinear filtering operations, which are performed by a convolutional autoencoder (CAE). The CAE provides the decomposition in space. 
Second, we predict the time evolution of the turbulent state in the latent space, which is performed by an echo state network (ESN). The ESN provides the decomposition in time. 
Third, by assembling the CAE and the ESN, we obtain an autonomous dynamical system: the convolutional autoncoder echo state network (CAE-ESN). This is the reduced-order model of the turbulent flow.
We test the CAE-ESN on a two-dimensional flow. 
We show that, after training, the CAE-ESN
(i) finds a latent-space representation of the turbulent flow that has less than 1\% of the degrees of freedom than the physical space;
(ii) time-accurately and statistically predicts the flow in both quasiperiodic and turbulent regimes; 
(iii) is robust for different flow regimes (Reynolds numbers); and 
(iv) takes less than 1\% of computational time to predict the turbulent flow than solving the governing equations. 
This work opens up new possibilities for  nonlinear decompositions and reduced-order modelling of turbulent flows from data. 

\end{abstract}


\section{Introduction}

In the past decades, large amount of data has been generated from experiments and numerical simulations of turbulent flows \citep[e.g.,][]{durasaimy2019}. 
To analyse high-dimensional data, low-order representations are typically sought   \citep[e.g.,][]{taira2017}. Reduced-order modelling consists of predicting the time evolution of high-dimensional systems with a low-order representation of the system. By predicting the system based on a (relatively) small number of degrees of freedom, reduced-order modelling significantly reduces the computational cost and provides insights into the physics of the system. In this paper, we propose a nonlinear decomposition of the turbulent state. We construct a reduced-order model by (i) inferring a lower-dimensional manifold, the \emph{latent space}, where the turbulent dynamics live; and (ii)  predicting the turbulent dynamics in the latent space.
%
To generate the latent space, techniques such as Proper Orthogonal Decomposition (POD) \citep{lumley1967structure} and Dynamic Mode Decomposition (DMD) \citep{schmid2010dynamic} are commonly used. They have been successfully applied in mutiple settings, such as extracting spatiotemporal features and controlling flowfields \citep[e.g.,][]{ROWLEY2004, Rowley2017}.
The downside of these methodologies is that they are linear approximators, which require a large number of modes to describe turbulent flowfields \citep[e.g.,][]{muralidhar_2019, alfonsi2007structure}.  To reduce the number of modes to accurately describe flowfields, \emph{nonlinear} mapping through machine learning has shown promising results in recent years \citep[e.g.,][]{agostini2020exploration,brunton2020machine}. 
A robust data-driven method that computes a low-dimensional representation of the data is the autoencoder \citep{kramer1991, goodfellow2016deep}. Autoencoders typically consist of a series of neural networks that map the original field to (and back from) the latent space. In fluids, \citet{MILANO2002} developed a feed-forward neural network autoencoder to investigate a turbulent channel flow. Since then, the advent of data-driven techniques tailored for the analysis of spatially varying data, such as convolutional neural networks (CNNs) \citep{lecun1998}, has greatly extended the applicability of autoencoders \citep{hinton2006}. For example, \citet{fukami_2019} employed CNNs to improve the resolution of sparse measurements of a turbulent flowfield, and \citet{murata2020} analysed the autoencoder modes in the laminar wake past a cylinder. 

Once the latent space is generated, we wish to predict the temporal dynamics within the latent space. To do so, one option is to project the governing equations onto the low-order space \citep{antoulas2005approximation}. This is a common method when the governing equations are known, but it becomes difficult to implement when the equations are not exactly known \citep{yu2019}.
A reduced-order modelling approach that does not require governing equations is data-driven, which is the focus of this paper. 
For forecasting temporal dynamics,  Recurrent Neural Networks (RNNs) \citep{rumelhart1986learning} are the state-of-the-art data-driven architectures  \citep[e.g.,][]{goodfellow2016deep,chattopadhyay2019data}. RNNs are designed to infer the correlation within data sequentially ordered in time via an internal hidden state, which is updated at each time step. In fluids, RNNs have been deployed for predicting flows past bluff bodies of different shapes \citep{Hasegawa_2020}, replicating the statistics of turbulence \citep{snrinivasan2020,nakamura2021convolutional}, and  controlling  gliding \citep{novati2019}, to name a few.
Among recurrent neural networks, reservoir computers in the form of Echo State Networks (ESNs) \citep{jaeger2004harnessing,maass2002real} are versatile architectures for the prediction of chaotic dynamics. ESNs are universal approximators under non-stringent assumptions of ergodicity \citep{GRIGORYEVA2018,hart2021echo}, which perform (at least) as well as  other architectures such as Long Short-Term Memory (LSTM) networks \citep{vlachas2020backpropagation}, but they are simpler to train.
ESNs are designed to be straightforward and computationally cheaper to train than other networks, but their performance is sensitive to the selection of hyperparameters, which need to be selected through ad-hoc algorithms \citep{racca2021robust}.
In fluid dynamics, echo state networks have been employed to (i) optimize ergodic averages in thermoacoustic oscillations \citep{huhn2021gradient}, (ii) predict extreme events in chaotic flows \citep{doan2021short, racca2022statistical} and control their occurrence \citep{racca2022data}, and (iii) infer the model error (i.e., bias) in data assimilation of thermoacoustics systems \citep{novoa2021real,novoainter}.

The objective of this work is threefold. 
First, we develop the Convolutional AutoEncoder Echo State Network (CAE-ESN) by combining convolutional autoencoders with echo state networks to predict turbulent flowfields.
Second, we time-accurately and statistically predict a turbulent flow at different Reynolds numbers. 
Third, we carry out a correlation analysis between the reconstruction error and the  temporal prediction of the CAE-ESN. 
%
The paper is organised as follows.
Section \ref{sec:data} presents the two dimensional turbulent flow and introduces the tools for nonlinear analysis. 
Section \ref{sec: ML} describes the CAE-ESN.
Section \ref{sec:dec-rec} analyses the reconstruction of the flowfield.  
Section \ref{sec:time-acc} analyses the time-accurate prediction of the flow and discusses the correlation between reconstruction error and temporal prediction of the system.
Section \ref{sec:statistics} analyses the prediction of the statistics of the flow and the correlation between time-accurate and statistical performance.
Finally, \S \ref{sec:conc} summarizes the results. 

\section{Kolmogorov flow}
\label{sec:data}

We consider the two-dimensional non-dimensionalised incompressible Navier–Stokes equations 
\begin{gather}
    \frac{\partial\boldsymbol{u}}{\partial t} + \boldsymbol{u}\cdot\bnabla\boldsymbol{u} = - \bnabla p + \frac{1}{\Rey}\Delta\boldsymbol{u} + \boldsymbol{f}, \label{eq:koleqs1} \\ 
    \bnabla\bcdot\boldsymbol{u} = 0, \label{eq:koleqs2}
\end{gather}
where $p$ is the pressure, $\boldsymbol{u}=(u_1,u_2)$ is the velocity and $\boldsymbol{x}=(x_1,x_2)$ are the spatial coordinates. The time-independent diverge-free forcing is $\boldsymbol{f}=\sin(n_fx_2)\boldsymbol{e}_1$, where $\boldsymbol{e}_1= (1,0)^T$ and $n_f$ is the wavenumber of the forcing. 
The Reynolds number is $\Rey=\sqrt{\chi}/\nu$, where $\chi$ is the amplitude of the forcing and $\nu$ is the kinematic viscosity \citep{chandler_2013}. We solve the flow in the doubly periodical domain $\boldsymbol{x} \in \mathbb{T}^2=[0,2\pi]\times[0,2\pi]$. For this choice of forcing and boundary conditions, the flow is typically referred to as the Kolmogorov flow \citep[e.g.,][]{platt1991investigation}. 
We integrate the equations using \href{https://github.com/MagriLab/KolSol}{KolSol}, which is a publicly available pseudospectral solver based on the Fourier-Galerkin approach described by \citet{canuto1988spectral}. 
The equations are solved in the Fourier domain with a 4-$th$ order explicit Runge-Kutta integration scheme with a timestep $dt=0.01$. The results are stored every $\delta t=0.1$. 
As suggested in~\citet{farazmand_2016_adjoint}, we select the number of Fourier modes from convergence tests on the kinetic energy spectra (see Supplementary Material). The solution in the Fourier domain is projected onto a $48\times48$ grid in the spatial domain with the inverse Fourier transform. The resulting 4608-dimensional velocity flowfield, $\boldsymbol{q}(t) \in \mathbb{R}^{48\times48\times2}$, is the flow state vector.

\subsection{Regimes}
\begin{figure}
    \centering
    \includegraphics[width=.85\textwidth]{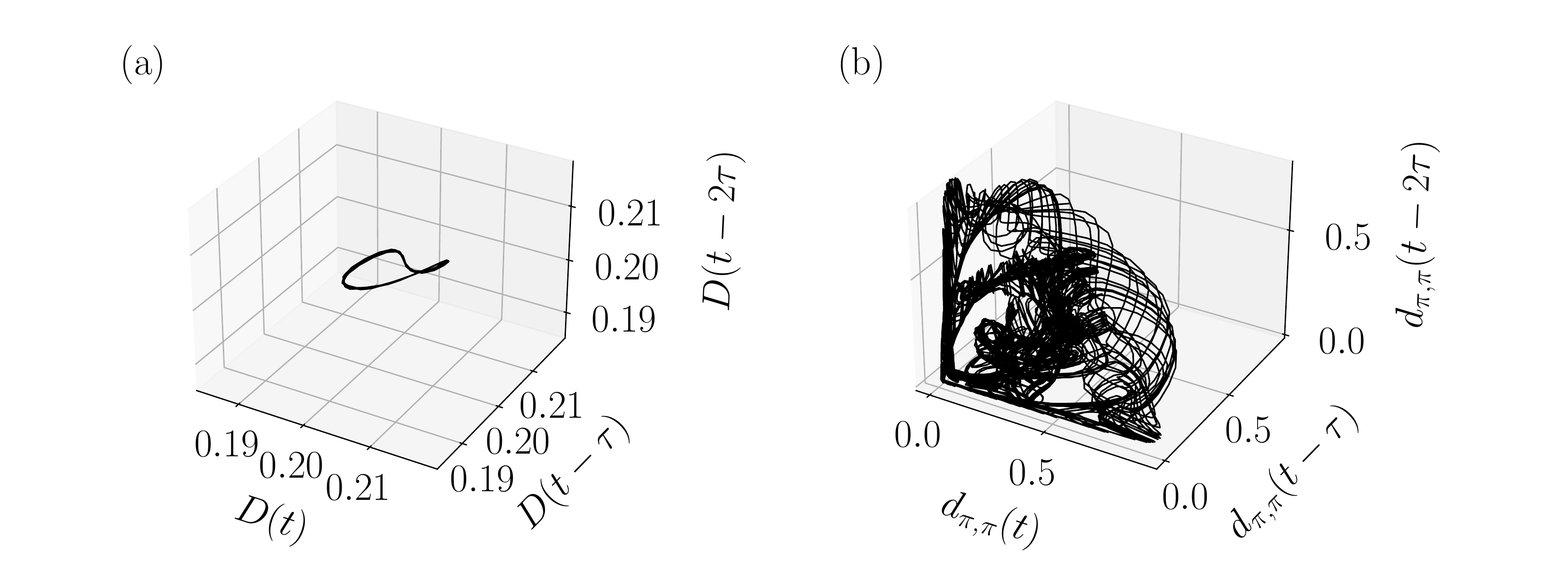}
    \includegraphics[width=.85\textwidth]{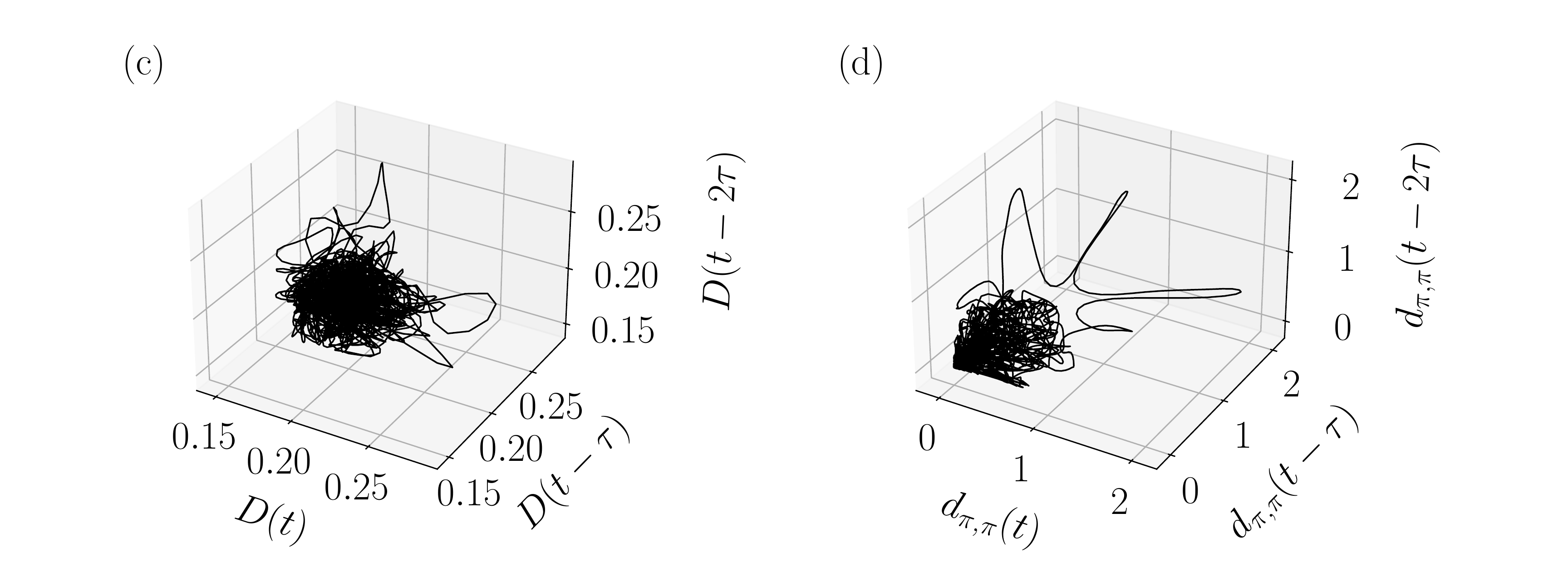}
    \includegraphics[width=.8\textwidth]{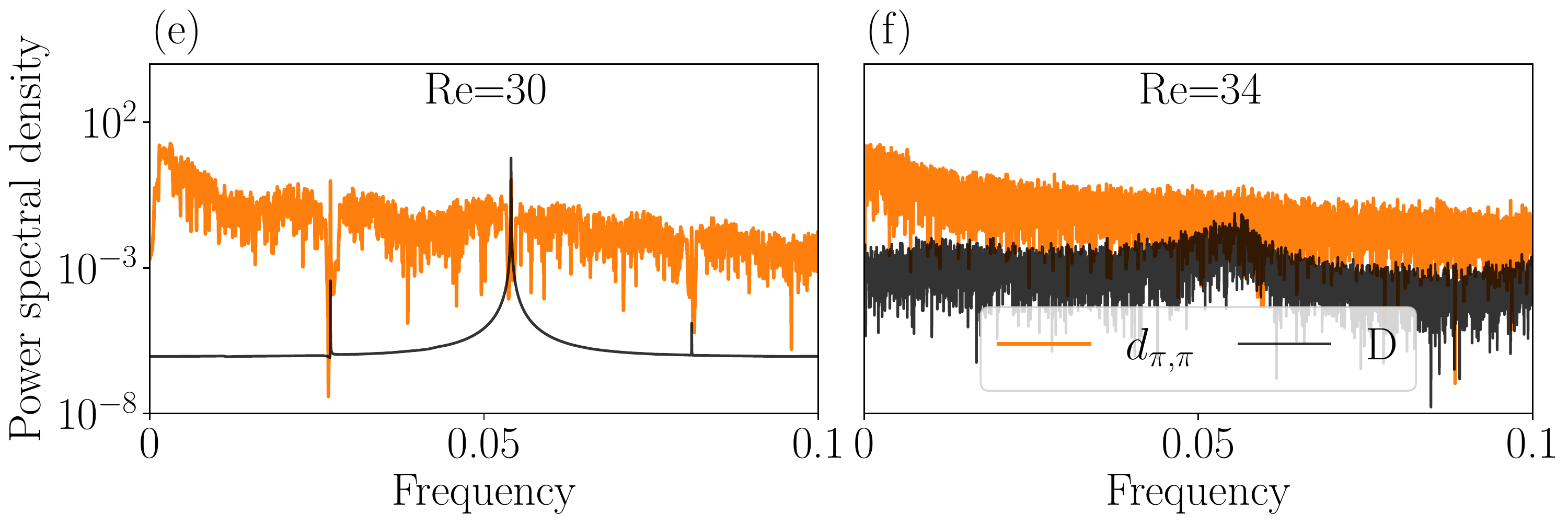}
    \caption{Phase plots of the global, $D$, and local, $d_{\pi,\pi}$, dissipation rates for (a)-(b) quasiperiodic regime of the flow state ($\Rey=30$), and (c)-(d) turbulent regime ($\Rey=34$). Power spectral density for (e) $\Rey=30$, and (f) $\Rey=34$.}
    \label{fig:attractors}
\end{figure}

\begin{figure}
    \centering
    \includegraphics[width=1.\textwidth]{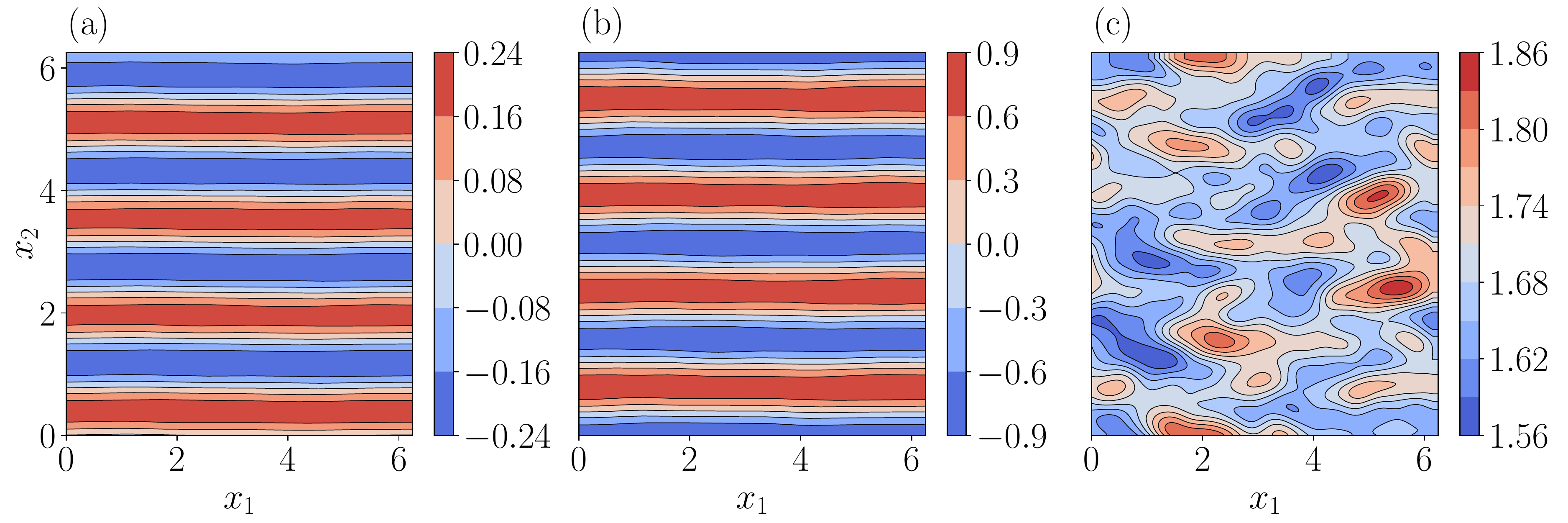}
    \caption{Flowfield of the time-averaged (a) velocity component along $x_1$, $\overline{u}_1$, (b) vorticity component, $\overline{w}$, and (c) enstrophy, $\overline{\omega}$, in the turbulent regime ($\Rey=34$). }
    \label{fig:average_flowfields}
\end{figure}

The Kolmogorov flow shows a variety of regimes that depend on the forcing wavenumber, $n_f$, and Reynolds number, $\Rey$ \citep{platt1991investigation}.
In this work, we analyse $n_f=4$ and $\Rey=\{30,34\}$, for which we observe  quasiperiodic and turbulent solutions,  respectively.
To globally characterize the flow, we compute the average dissipation rate, $D$, per unit volume
\begin{gather}
   \qquad D(t) = \frac{1}{(2\pi)^2} \int_0^{2\pi} \int_0^{2\pi} d_{x_1,x_2}(t) \; \mathrm{d}x_1\mathrm{d}x_2, \qquad
    d_{x_1,x_2}(t) = \frac{1}{\Rey}|| \bnabla \boldsymbol{u} (x_1,x_2) ||^2,
    \label{eq:dissipation}
\end{gather}
where $d_{x_1,x_2}(t)$ is the local dissipation rate and $||\cdot||$ is the $L_2$ norm.
The dissipation rate has been  employed in the literature to analyse the Kolmogorov flow \citep{chandler_2013, farazmand_2016_adjoint}.
%
We characterize the solutions in Figure~\ref{fig:attractors}.
The phase plots show
 the trajectories of the average and local dissipation rates with the optimal time delay given by the first minimum of the average mutual information \citep{kantz2004nonlinear}. In the first regime ($Re=30$), the average dissipation rate is a limit cycle (Figure \ref{fig:attractors}a), whilst the local dissipation rate has a toroidal structure (Figure \ref{fig:attractors}b), which indicates quasiperiodic variations in the flow state.
The average dissipation rate is periodic, despite the flow state being quasiperiodic, because temporal frequencies are filtered out when averaging in space (Figure \ref{fig:attractors}e). 
In the second regime ($Re=34$), the solution is turbulent for both global and local quantities (Figure~\ref{fig:attractors}c,d,f). The time average, $\overline{(\cdot)}$, over 300,000 time units (i.e., $3\times10^7$ integration steps) of the first velocity component, $u_1$, and of the vorticity component, $w$,
are plotted in Figure~\ref{fig:average_flowfields}. The velocity component (Figure~\ref{fig:average_flowfields}a), is on average aligned with the forcing, $\sin(n_fx_2)$, but it has a smaller magnitude than the unstable laminar state, $u_1 = \Rey/n_f^2\sin(n_fx_2)$ \citep{chandler_2013}. Likewise, the time-averaged vorticity component shows a similar spatial structure (Figure~\ref{fig:average_flowfields}b). This is because its computation is dominated by the term $\partial u_1/ \partial x_2$, which causes the spatial phase shift with respect to $u_1$. The local enstrophy, $\omega=\frac{1}{2}w^2$, shows rich spatial structures, which are characteristic of turbulent dynamics (Figure~\ref{fig:average_flowfields}c). 

\subsection{Lyapunov exponents and attractor dimension}

\label{sec:KY}

As explained in \S\ref{sec:dec-rec}, in order to create a reduced-order model, we need to select the number of degrees of freedom of the latent space. 
We observe that at a statistically stationary regime, the latent space should be at least as large as the turbulent attractor. Therefore,  we propose using the turbulent attractor's dimension as a lower bound for the  latent space dimension. In chaotic (turbulent) systems, the dynamics are predictable only for finite times because infinitesimal errors increase in time with an average exponential rate given by the (positive) largest Lyapunov exponent, $\Lambda_1$ \citep[e.g.,][]{boffetta2002predictability}. By  definition, the inverse of the largest Lyapunov exponent provides a timescale for assessing the predictability of chaotic systems, which is referred to as the Lyapunov time, $1\mathrm{LT}=\Lambda_1^{-1}$. To quantitatively assess time accuracy (\S \ref{sec:time-acc}), we normalize the time by the Lyapunov time.
In addition to being unpredictable, chaotic systems are dissipative, which means that the solution converges to a limited region of the phase space, i.e. the attractor.  The attractor has typically a significantly smaller number of degrees of freedom than the original system \citep{eckmann1985ergodic}. 
An upper bound on the number of degrees of freedom of the attractor, i.e. its dimension, can be estimated via the Kaplan-Yorke dimension \citep{kaplan1979chaotic}
\begin{equation}
    N_{KY} = j +  \frac{\sum_{i=1}^j\Lambda_i}{|\Lambda_{j+1}|},
\end{equation}
where $\Lambda_i$ are the $j$ largest Lyapunov exponents for which $\sum_{i=1}^j\Lambda_i \geq 0$.
Physically, the $m$ largest Lyapunov exponents are the average exponential contraction/expansion rates of an $m$-dimensional infinitesimal volume of the phase space moving along the attractor \citep[e.g.,][]{boffetta2002predictability}.
To obtain the $m$ largest Lyapunov exponents, we compute the evolution of $m$ random perturbations around a trajectory that spans the attractor. The space spanned by the $m$ perturbations approximates an $m$-dimensional subspace of the tangent space. Because errors grow exponentially in time, the evolution of the perturbations is exponentially unstable and the direct computation of the Lyapunov exponents numerically overflows. To overcome this issue, we periodically orthonormalize the perturbations, following the algorithm of \citet{benettin1980lyapunov} (see Supplementary material). In so doing, we find the quasiperiodic attractor to be 4-dimensional
and the chaotic attractor to be 9.5-dimensional. Thus, both attractors have approximately three order of magnitude fewer degrees of freedom than the flow state (which has 4608 degrees of freedom, see \S \ref{sec:data}). We will take advantage of these estimates in \S\ref{sec:dec-rec}-\ref{sec:time-acc}. The leading Lyapunov exponent in the chaotic case is $\Lambda_1=0.065$, therefore, the Lyapunov time is $1\mathrm{LT}=0.065^{-1}\approx 15.4$. 

\section{Convolutional AutoEncoder Echo State Network (CAE-ESN)}

\label{sec: ML}

In order to decompose high-dimensional turbulent flows into a lower-order representation, a latent space of the physical dynamics is computed. 
For time prediction, the dynamics are mapped onto the latent space on which their evolution can be predicted at a lower computational cost than the original problem.
In this work, we generate the low-dimensional space using a Convolutional AutoEncoder (CAE) \citep{hinton2006}, which offers a nonlinear reduced-order representation of the flow. 
By projecting the physical dynamics onto the latent space, we obtain a low dimensional time-series, whose dynamics are predicted by an Echo State Network (ESN) \citep{jaeger2004harnessing}. 

\subsection{Convolutional autoencoder}

\begin{figure}
    \centering
    \includegraphics[width=1.\textwidth]{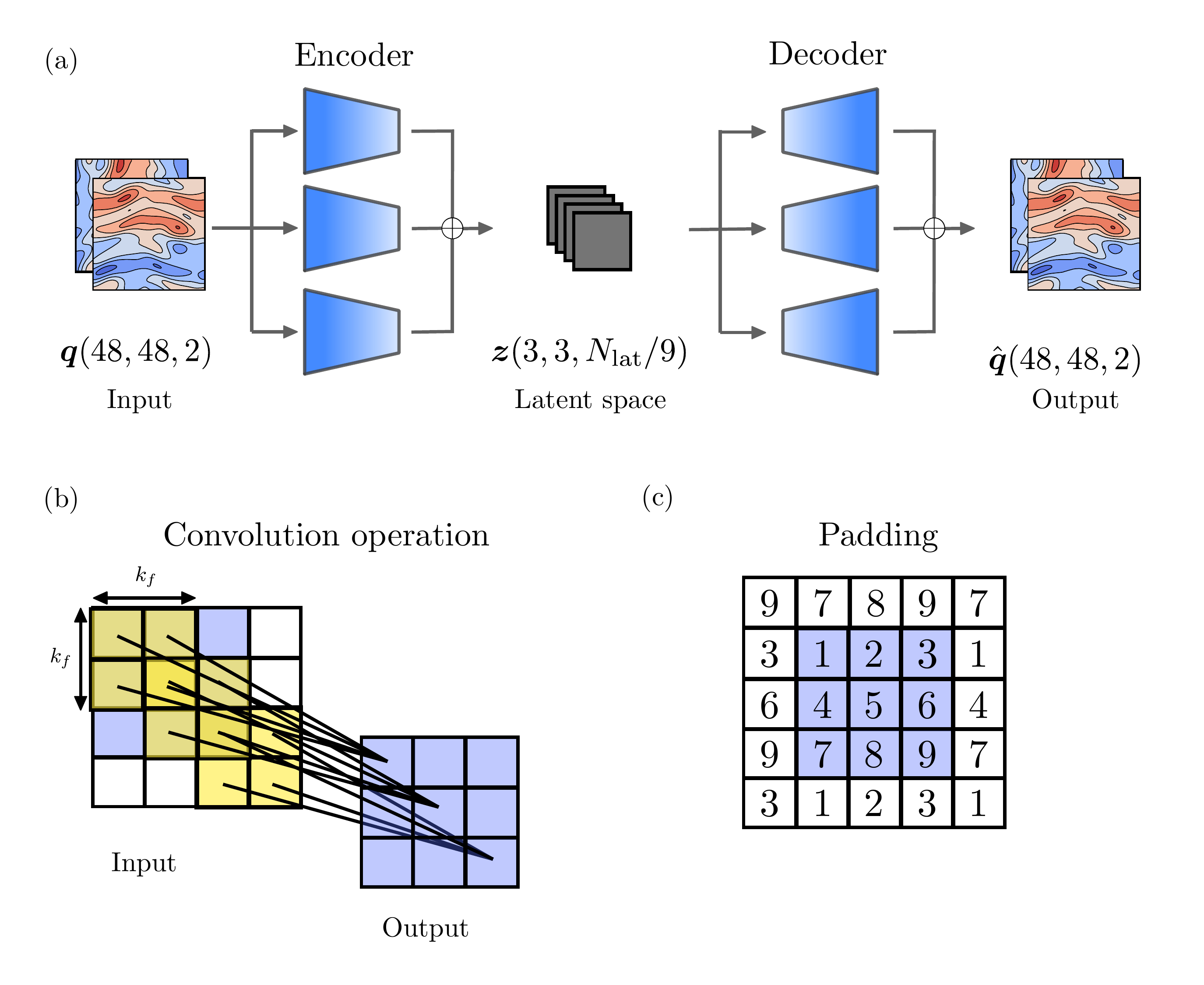}
    \caption{(a) Schematic representation of the multiscale autoencoder, (b) convolution operation for filter size $(2\times2\times1)$, stride=1 and padding=1, and (c) periodic padding. In (b) and (c), blue and white squares indicate the input and the padding, respectively. The numbers in (c) indicate pictorially the values of the flowfield, which can be interpreted as pixel values.}
    \label{fig:multiscale_autoencoder}
\end{figure}

\label{sec:CNN}

The autoencoder consists of an encoder and a decoder (Figure~\ref{fig:multiscale_autoencoder}a). The encoder, $\boldsymbol{g}(\cdot)$,  maps the high-dimensional physical state, $\boldsymbol{q} \in \mathbb{R}^{N_{\mathrm{phys}}}$, into the low-dimensional latent state, $\boldsymbol{z} \in \mathbb{R}^{N_{\mathrm{lat}}}$ with $N_{\mathrm{lat}} \ll N_{\mathrm{phys}}$;  whereas the decoder, $\boldsymbol{f}(\cdot)$, maps the latent state back into the physical space with the following goal
\begin{equation}
\hat{\boldsymbol{q}} \simeq \boldsymbol{q}, 
 \quad\textrm{where}\quad
\hat{\boldsymbol{q}} = \boldsymbol{f}(\boldsymbol{z}), \quad \boldsymbol{z} = \boldsymbol{g}(\boldsymbol{q}), 
\end{equation}
where $\hat{\boldsymbol{q}} \in \mathbb{R}^{N_{\mathrm{phys}}}$ is the reconstructed state.
We use Convolutional Neural Networks (CNNs) \citep{lecun1998} as the building blocks of the autoencoder. In CNNs, a filter of size $k_f\times k_f\times d_f$, slides through the input, $\boldsymbol{y}_1 \in \mathbb{R}^{N_{1x}\times N_{1y}\times N_{1z}}$, with stride, $s$, so that the output, $\boldsymbol{y}_2 \in \mathbb{R}^{N_{2x}\times N_{2y}\times N_{2z}}$, of the CNN layer is 
\begin{equation}
    y_{2_{ijm}} = \mathrm{func} \left(\sum_{l_1=1}^{k_f} \sum_{l_2=1}^{k_f} \sum_{k=1}^{N_{1z}} y_{1_{s(i-1)+l_1,s(j-1)+l_2,k}}W_{l_1l_2km} + b_{m} \right), 
\end{equation}
\noindent where $\mathrm{func}$ is the nonlinear activation function, and
$W_{l_1l_2km}$ and $b_{m}$ are the weights and bias of the filter, which are computed by training the network. We choose $\mathrm{func}=\tanh$ following \cite{murata2020}. A visual representation of the convolution operation is shown in Figure~\ref{fig:multiscale_autoencoder}b.
The size of the output is a function of the input size, the kernel size, the stride and the artificially added entries around the input (i.e., the padding). By selecting periodic padding, we enforce the boundary conditions of the flow (Figure~\ref{fig:multiscale_autoencoder}c).
The convolution operation filters the input by analyzing patches of size $k_f \times k_f$. In doing so, CNNs take into account the spatial structure of the input and learn localized structures in the flow. Moreover, the filter has the same weights as it slides through the input (parameter sharing), which allows us to create models with significantly less weights than fully connected layers. Because the filter provides the mathematical relationship between nearby points in the flowfield, parameter sharing is physically consistent with the governing equations of the flow, which are invariant to translation. 

The encoder consists of a series of padding and convolutional layers. At each stage, we (i) apply periodic padding, (ii) perform a convolution with stride equal to 2 to half the spatial dimensions and increase the depth of the output and (iii) perform a convolution with stride equal to 1 to keep the same dimensions and increase the representation capability of the network. The final convolutional layer has a varying depth, which depends on the size of the latent space. 
The decoder consists of a series of padding, transpose convolutional and center crop layers. An initial periodic padding enlarges the latent space input through the periodic boundary conditions. The size of the padding is chosen so that the spatial size of the output after the last transpose convolutional layer is larger than the physical space. In the next layers, we (i) increase the spatial dimensions of the input through the transpose convolution with stride equal to 2 and (ii) perform a convolution with stride equal to 1 to keep the same dimensions and increase the representation capability of the network. The transpose convolution is the inverse operation of the convolution shown in Figure~\ref{fig:multiscale_autoencoder}b, in which the roles of the input and the output are inverted \citep{zeiler2010}.  The center crop layer eliminates the outer borders of the picture after the transpose convolution, in a process opposite to padding, which is needed to match the spatial size of the output of the decoder with the spatial size of the input of the encoder. A final convolutional layer with a linear activation function sets the depth of the output of the decoder to be equal to the input of the encoder. We use the linear activation to match the amplitude of the inputs to the encoder. 
The encoder and decoder have similar number of trainable parameters
(more details in Appendix \ref{sec:tab_layers}). \\

In this study, we use a multi-scale autoencoder \citep{du2018single}, which has been successfully employed to generate latent spaces of turbulent flowfields \citep{nakamura2021convolutional}. The multi-scale autoencoder employs three parallel encoders and decoders, which have different spatial sizes of the filter (Figure~\ref{fig:multiscale_autoencoder}a). The size of the three filters are $3\times3$, $5\times5$ and $7\times7$, respectively.
By employing different filters, the multi-scale architecture learns spatial structures of different sizes, which are characteristic features of turbulent flows.
We train the autoencoder by minimising the Mean Square Error (MSE) between the outputs and the inputs
\begin{equation}
    \mathcal{L} = \sum^{N_t}_{i=1}
    \frac{1}{ N_tN_{\mathrm{phys}}}
    \left\vert\left\vert\hat{\boldsymbol{q}}(t_i) - \boldsymbol{q}(t_i)\right\vert\right\vert^2,
\label{eq:loss}
\end{equation}
where $N_t$ is the number of training snapshots. To train the autoencoder, we use a dataset of 30000 time units (generated by integrating Equations~\eqref{eq:koleqs1}-\eqref{eq:koleqs2}), which we sample with timestep $\delta t_{\mathrm{CNN}} = 1$. Specifically, we use 25000 time units for training and 5000 for validation. We divide the training data in minibatches of 50 snapshots each, where every snapshot is $500\delta t_{\mathrm{CNN}}$ from the previous input of the minibatch. The weights are initialized following \citet{glorot2010}, and the minimisation is performed by stochastic gradient descent with the AMSgrad variant of the Adam algorithm \citep{kingma2014adam, reddi2019convergence} with adaptive learning rate. The autoencoder is implemented in Tensorflow \citep{tensorflow2015-whitepaper}.


\begin{figure}
\centering
\includegraphics[width=1.\textwidth]{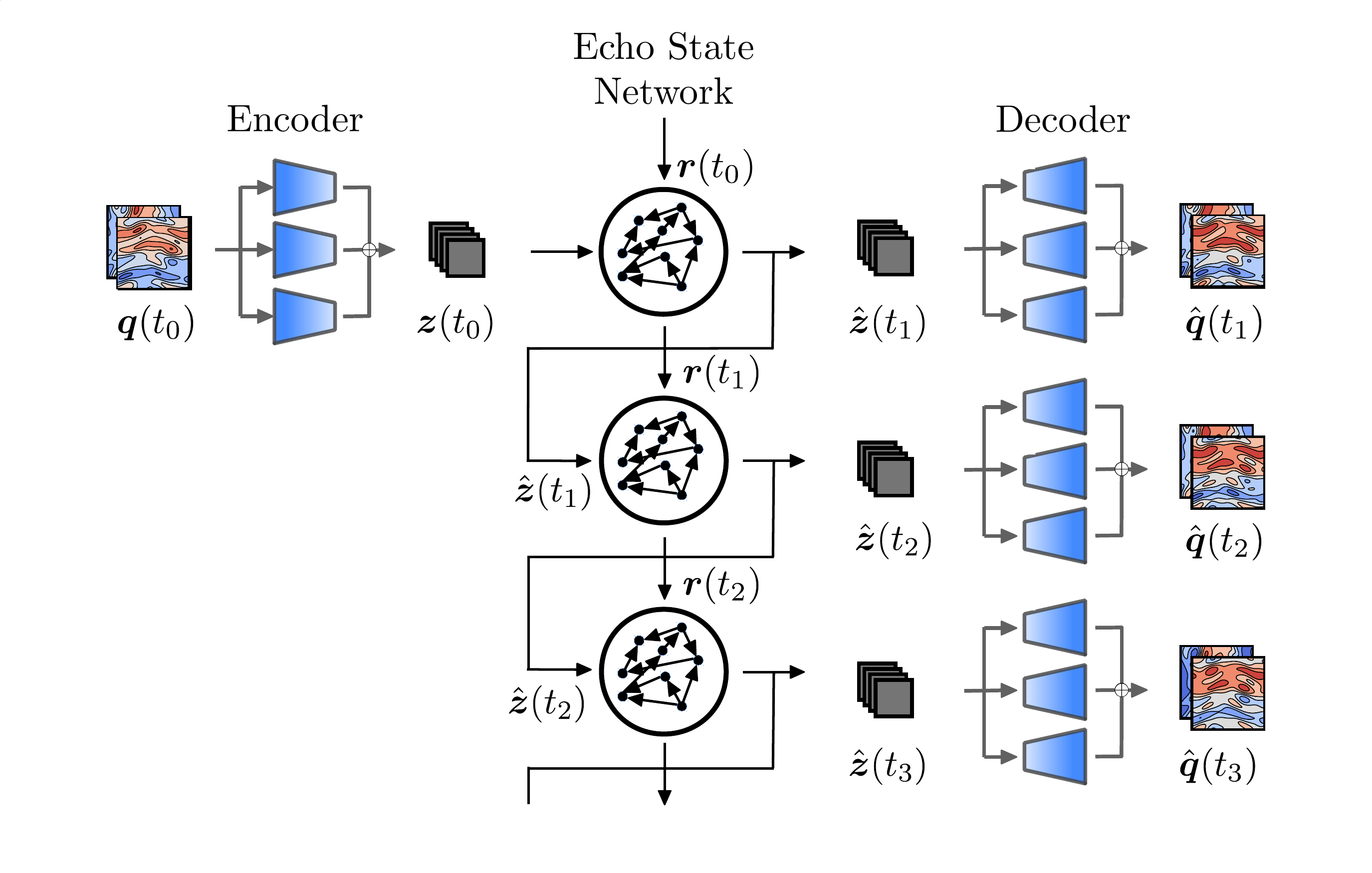}
\caption{Schematic representation of the closed-loop evolution of the echo state network in the latent space, which is decompressed by the decoder.}
\label{fig:ESN}
\end{figure}

\subsection{Echo state networks}
\label{sec:ESN}

Recurrent neural networks keep memory of the state of the physical system by updating an internal hidden state. Because of the long-lasting time dependencies of the hidden state, however, training RNNs with backpropagation through time is notoriously difficult \citep{werbos1990backpropagation}. Echo state networks overcome this issue by nonlinearly expanding the inputs into a higher-dimensional system, the reservoir, which acts as the memory of the system \citep{lukovsevivcius2012practical}. The output of the network is a linear combination of the reservoir's dynamics, whose weights are the only trainable parameters of the system. Thanks to this  architecture, training ESNs consists of a straightforward linear regression problem, which bypasses backpropagation through time.\\

As shown in Figure~\ref{fig:ESN}, in an echo state network, at any time $t_i$, (i) the latent state input, $\boldsymbol{z}(t_i)$, is mapped into the reservoir state, by the input matrix, $\mathsfbi{W}_{\mathrm{in}} \in \mathbb{R}^{N_r\times (N_{\mathrm{lat}}+1)}$, where $N_r > N_{\mathrm{lat}}$; (ii) the reservoir state, $\boldsymbol{r} \in \mathbb{R}^{N_r}$, is updated at each time iteration as a function of the current input and its previous value, and (iii) the updated reservoir is used to compute the output, which is the predicted latent state at the next timestep, $\hat{\boldsymbol{z}}(t_{i+1})$. This process yields the discrete dynamical equations that govern the echo state network's evolution \citep{lukovsevivcius2012practical}
\begin{gather}
\label{state_step}
        \boldsymbol{r}(t_{i+1}) = \tanh\left(\mathsfbi{W}_{\mathrm{in}}[\Tilde{\boldsymbol{z}}(t_i);b_\mathrm{in}]+\mathsfbi{W}\boldsymbol{r}(t_i)\right), \nonumber \\
        \hat{\boldsymbol{z}}(t_{i+1}) = [\boldsymbol{r}(t_{i+1});1]^T\mathsfbi{W}_{\mathrm{out}}, 
\end{gather}
where $\Tilde{(\cdot)}$ indicates that each component is normalized by its range, $\mathsfbi{W} \in \mathbb{R}^{N_r\times N_r}$ is the state matrix, $b_{\mathrm{in}}$ is the input bias and $\mathsfbi{W}_{\mathrm{out}} \in \mathbb{R}^{(N_{r}+1)\times N_{\mathrm{lat}}}$ is the output matrix. 
The matrices $\mathsfbi{W}_{\mathrm{in}}$ and $\mathsfbi{W}$ are (pseudo)randomly generated and fixed, whilst the weights of the output matrix, $\mathsfbi{W}_{\mathrm{out}}$, are computed by training the network. The input matrix,
$\mathsfbi{W}_{\mathrm{in}}$, has only one element different from zero per row, which is sampled from a uniform distribution in $[-\sigma_{\mathrm{in}},\sigma_{\mathrm{in}}]$, where $\sigma_{\mathrm{in}}$ is the input scaling. The state matrix, $\mathsfbi{W}$, is an Erdős-Renyi matrix with average connectivity, $d=3$, in which each neuron (each row of $\mathsfbi{W}$) has on average only $d$ connections (non-zero elements), which are obtained by sampling from a uniform distribution in $[-1,1]$; the entire matrix is then scaled by a multiplication factor to set the spectral radius, $\rho$. The value of the connectivity is kept small to speed up the computation of $\mathsfbi{W}\boldsymbol{r}(t_i)$, which, thanks to the sparseness of $\mathsfbi{W}$, consists of only $N_rd$ operations.
The bias in the inputs and outputs layers are added to break the inherent symmetry of the basic ESN architecture \citep{lu2017reservoir}.  The input bias, $b_{\mathrm{in}}=0.1$ is selected for it to have the same order of magnitude of the normalized inputs, $\hat{\boldsymbol{z}}$, whilst  the output bias is determined by training $\mathsfbi{W}_{\mathrm{out}}$.

The ESN can be run either in open-loop or closed-loop. 
In the open-loop configuration, we feed the data as the input at each time step to compute the reservoir dynamics, $\boldsymbol{r}(t_i)$. We use the open-loop configuration for washout and training.
Washout is the initial transient of the network, during which we do not compute the output, $\hat{\boldsymbol{z}}(t_{i+1})$.
The purpose of washout is for the reservoir state to 
become (i) up-to-date with respect to the current state of the system, and (ii) independent of the arbitrarily chosen initial condition, $\boldsymbol{r}(t_0) = {0}$ (echo state property). 
After washout, we train the output matrix, $\mathsfbi{W}_{\mathrm{out}}$, by minimizing the mean square error between the outputs and the data over the training set.
Training the network on $N_{\mathrm{tr}} + 1$ snapshots consists of solving the linear system (ridge regression)
\begin{equation}
\label{RidgeReg}
    (\mathsfbi{R}\mathsfbi{R}^T + \beta \mathsfbi{I})\mathsfbi{W}_{\mathrm{out}} = \mathsfbi{R} \mathsfbi{Z}_{\mathrm{d}}^T,
\end{equation}
\noindent where $\mathsfbi{R}\in\mathbb{R}^{(N_r+1)\times N_{\mathrm{tr}}}$ and $\mathsfbi{Z}_{\mathrm{d}}\in\mathbb{R}^{N_{\mathrm{lat}}\times N_{\mathrm{tr}}}$ are the horizontal concatenations of the reservoir states with bias, $[\boldsymbol{r};1]$, and of the output data, respectively; 
$\mathsfbi{I}$ is the identity matrix and $\beta$ is the Tikhonov regularization parameter \citep{tikhonov2013numerical}. 
In the closed-loop configuration (Figure~\ref{fig:ESN}), starting from an initial data point as an input and an initial reservoir state obtained through washout, the output, $\hat{\boldsymbol{z}}$, is fed back to the network as an input for the next time step prediction. 
In doing so, the network is able to autonomously evolve in the future. 
After training, the closed-loop configuration is deployed for validation and test on unseen dynamics.

\subsubsection{Validation}

\label{sec:val}

During validation, we use part of the data to select the hyperparameters of the network by minimizing the error of the prediction with respect to the data. 
In this work, we optimize the input scaling, $\sigma_{\mathrm{in}}$, spectral radius, $\rho$, and Tikhonov parameter, $\beta$, which are the
key hyperparameters for the performance of the network \citep{lukovsevivcius2012practical}. 
We use a Bayesian optimization to select $\sigma_{\mathrm{in}}$ and $\rho$, and perform a grid search within  $[\sigma_{\mathrm{in}},\rho]$ to select $\beta$ \citep{racca2021robust, huhn2021gradient}. The range of the hyperparameters vary as a function of the testcase (see Supplementary material). In addition, we add to the training inputs, $\boldsymbol{z}_{tr}$, gaussian noise, $\mathcal{N}$, with a zero mean and standard deviation, such that $z_{tr_i} = z_i + \mathcal{N}(0,k_z\sigma(z_i))$, where $\sigma(\cdot)$ is the standard deviation and $k_z$ is a tunable parameter (Appendix \ref{app:esn_val}). Adding noise to the data improves the ESN forecasting of  chaotic dynamics because the network explores more regions around the attractor, thereby becoming more robust to errors  \citep{lukovsevivcius2012practical, vlachas2020backpropagation, racca2022data}.
To select the hyperparmeters, we employ the Recycle Validation (RV) \citep{racca2021robust}. The recycle validation is a tailored validation strategy for the prediction of dynamical systems with Recurrent Neural Networks, which has been shown to outperform other validation strategies, such as the single shot validation, in the prediction of chaotic flows (Appendix \ref{app:esn_val}). In the RV, the network is trained only once on the entire dataset, and validation is performed on multiple intervals already used for training. This is possible because recurrent neural networks operate in two configurations (open-loop and closed-loop), which means that the networks can be validated in closed-loop on data used for training in open-loop. 


\section{Spatial reconstruction}
\label{sec:dec-rec}

We analyse the ability of the autoencoder to create a reduced-order representation of the flowfield, i.e., the latent space. The focus is on the spatial reconstruction of the flow. Prediction in time is discussed in \S\ref{sec:time-acc}-\ref{sec:statistics}. 

\subsection{Reconstruction error}

The autoencoder maps the flowfield onto the latent space and then reconstructs the flowfield based on the information contained in the latent space variables. The reconstructed flowfield (the output) is compared to the original flowfield (the input). The difference between the two flowfields is the reconstruction error, which we quantify with the Normalized Root Mean Square Error (NRMSE)
\begin{equation}
    \mathrm{NRMSE}(\boldsymbol{q}) = \sqrt{\frac{\sum_i^{N_{\mathrm{phys}}} \frac{1}{N_{\mathrm{phys}}} (\hat{q}_i-q_{i})^2 }{\sum_i^{N_{\mathrm{phys}}} \frac{1}{N_{\mathrm{phys}}}\sigma(q_i)^2}},
    \label{eq:NRMSE}
\end{equation}
where $i$ indicates the $i$th component of the physical flowfield, $\boldsymbol{q}$, and the reconstructed flowfield, $\hat{\boldsymbol{q}}$, and $\sigma(\cdot)$ is the standard deviation. We compare the results for different sizes of the latent space with the reconstruction obtained by Proper Orthogonal Decomposition (POD) \citep{lumley1967structure}, also known as Principal Component Analysis \citep{pearson1901liii}. In  POD, the $N_{\mathrm{lat}}$-dimensional orthogonal basis that spans the reduced-order space is given by the eigenvectors $\boldsymbol{\Phi}^{(\mathrm{phys})}_i$ associated with the largest $N_{\mathrm{lat}}$ eigenvalues of the covariance matrix, $\mathsfbi{C}=\frac{1}{N_t-1}\mathsfbi{Q_d}^T\mathsfbi{Q_d}$, where $\mathsfbi{Q_d} 
$ 
is the vertical concatenation of the $N_t$ flow snapshots available during training, from which the mean has been subtracted. Both POD and the autoencoder minimize the same loss function \eqref{eq:loss}. However,  POD provides the optimal subspace onto which  linear projections of the data preserve its energy. On the other hand, the autoencoder provides a nonlinear mapping of the data, which is optimized to preserve its energy. (In the limit of linear activation functions, autoencoders  perform similarly to POD  \citep{BALDI_1989, MILANO2002, murata2020}.) The size of the latent space of the autoencoder is selected to be larger than the Kaplan-Yorke dimension, which is $N_{KY} = \{4, 9.5\}$ for the quasiperiodic and turbulent testcases, respectively (\S \ref{sec:KY}). 
We do so to account for the (nonlinear) approximation  introduced by the CAE-ESN, and numerical errors in the optimisation of the architecture. 

\begin{figure}
    \centering
    \includegraphics[width=1.\textwidth]{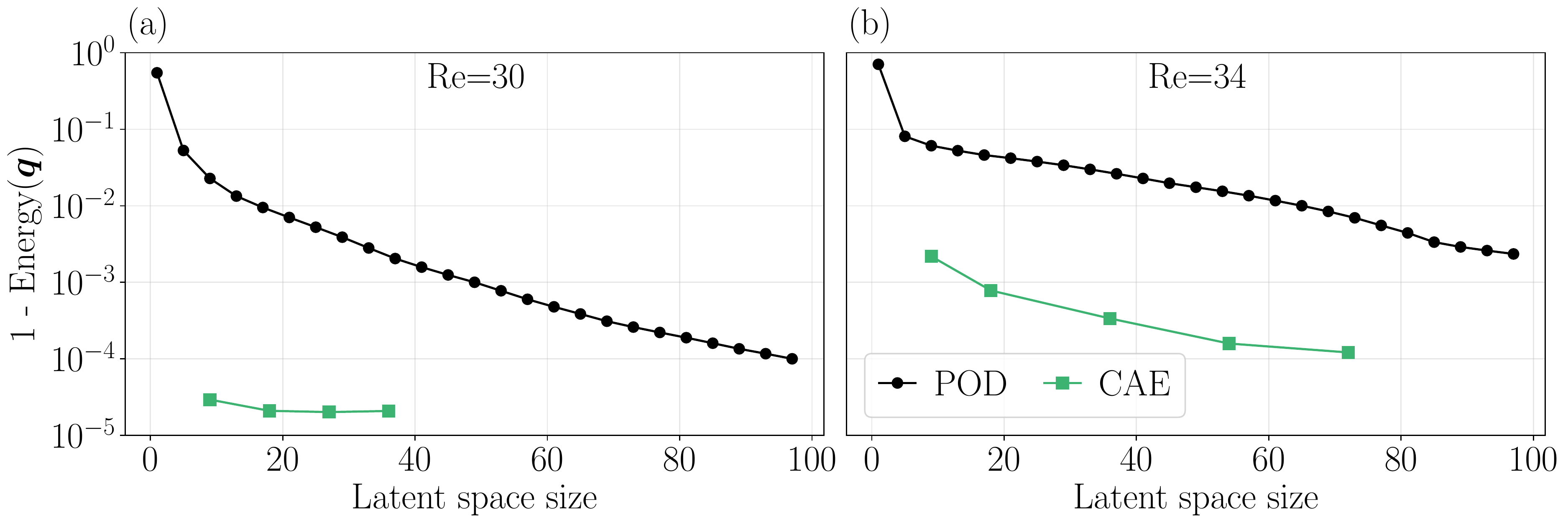}
    \caption{Reconstruction error in the test set as a function of the latent space size in (a) the quasiperiodic and (b) chaotic cases for POD and the convolutional autoencoder (CAE).}
    \label{fig:Rec}
\end{figure}

Figure~\ref{fig:Rec} shows the reconstruction error over 600000 snapshots in the test set for  POD and the autoencoder, in which we plot the energy (variance) captured by the modes
\begin{equation}
    \label{eq:energy}
    \mathrm{Energy}=1-\overline{\mathrm{NRMSE}}^2,
\end{equation}
where $\overline{\mathrm{NRMSE}}$ is the time-averaged NRMSE.
The rich spatial complexity of the turbulent flow (Figure~\ref{fig:Rec}b), with respect to the quasiperiodic solution (Figure~\ref{fig:Rec}a), is  apparent from the magnitude and slope of the reconstruction error as a function of the latent space dimension.
In the quasiperiodic case, the error is at least one order of magnitude smaller than the turbulent case for the same number of modes, showing that less modes are needed to characterize the flowfield. 
In both cases, the autoencoder is able to accurately reconstruct the flowfield. For example, in the quasiperiodic and turbulent cases the nine-dimensional autoencoder latent space captures $99.997\%$ and $99.78\%$ of the energy, respectively. The nine-dimensional autoencoder latent space provides a better reconstruction than 100 POD modes for the same cases. Overall, the autoencoder provides an error at least two orders of magnitude smaller than POD for the same size of the latent space. These results show that the nonlinear compression of the autoencoder  outperforms the optimal linear compression of POD.

\subsection{Autoencoder principal directions and POD modes}

\begin{figure}
    \centering
    \includegraphics[width=1.\textwidth]{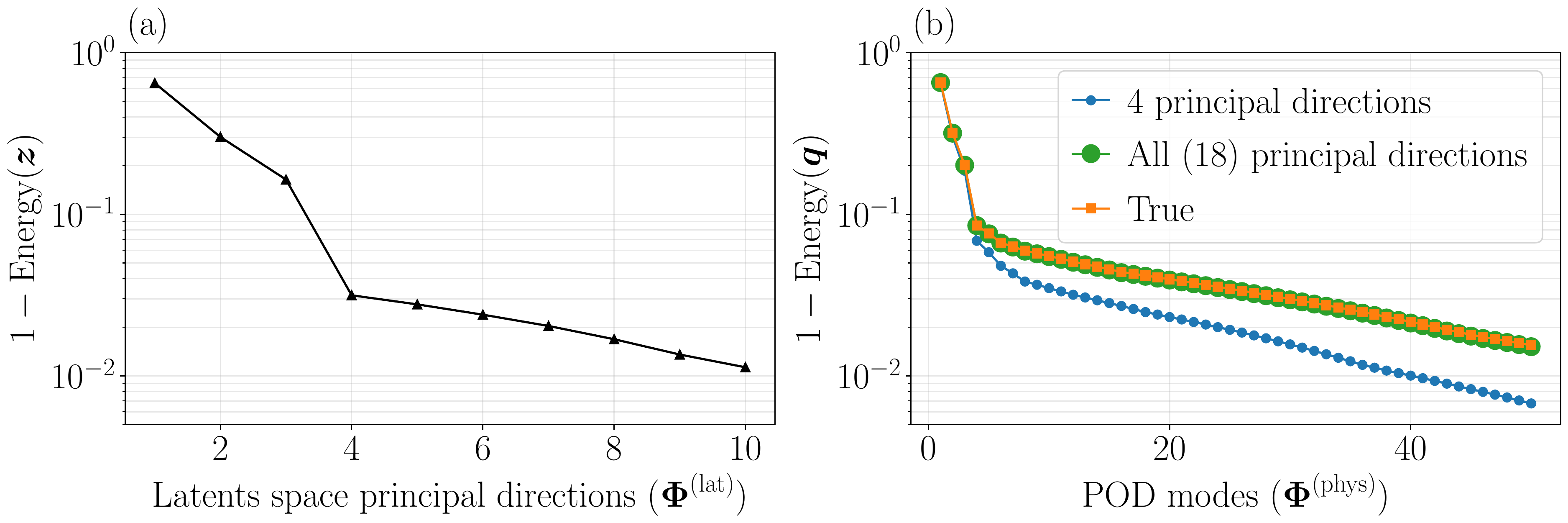}
    \caption{(a) Energy as a function of the latent space principal directions and (b) as function of the POD modes in the physical space for the field reconstructed using four and eighteen (all) latent space principal direction, and data (True).}
    \label{fig:lat_space_pod}
\end{figure}

\begin{figure}
    \centering
    \includegraphics[width=.73\textwidth]{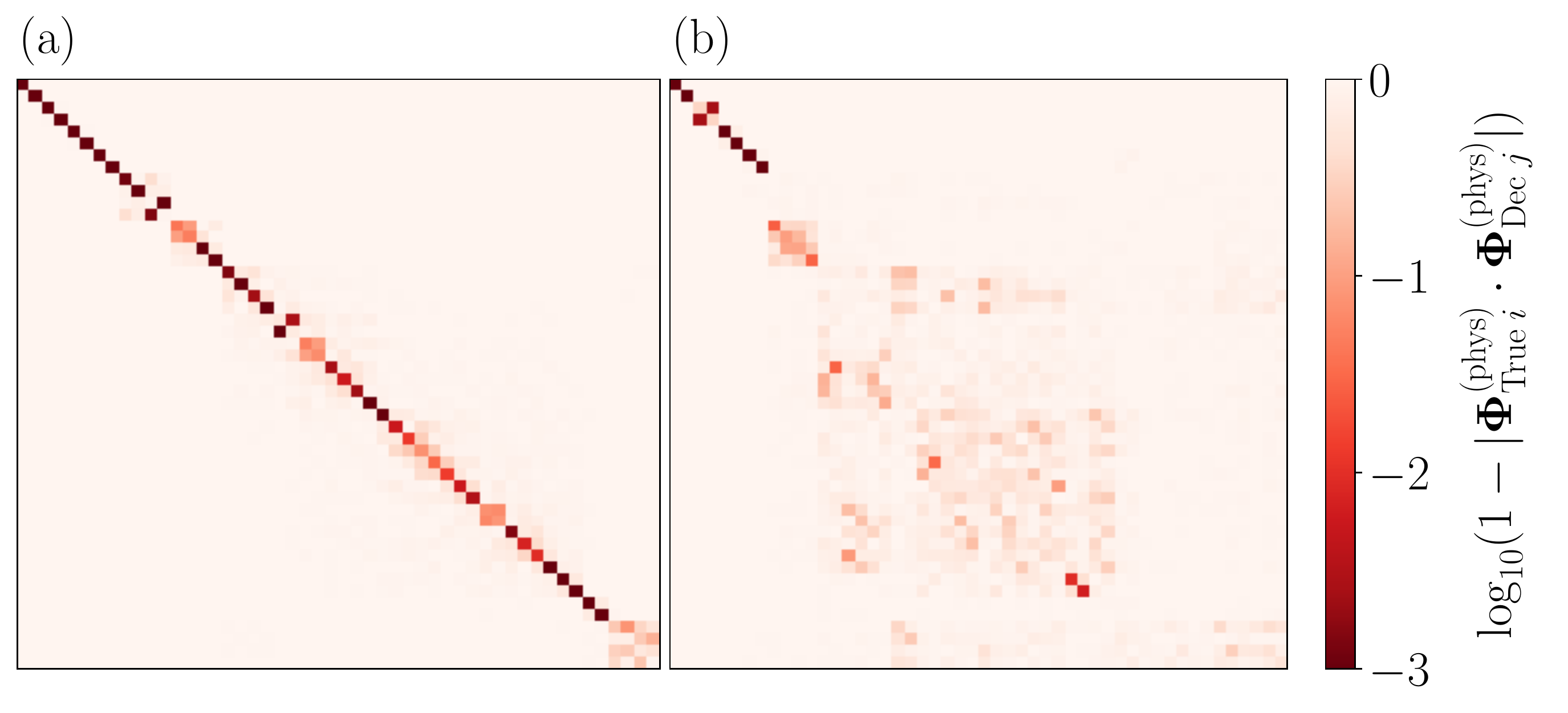}
    \caption{Scalar product of the POD modes of the data with (a) the POD modes of the decoded flowfield obtained from all the principal directions and (b) the POD modes of the decoded flowfield obtained from four principal directions.}
    \label{fig:modes_pod}
\end{figure}

We compare the POD modes provided by the autoencoder reconstruction, $\boldsymbol{\Phi}^{(\mathrm{phys})}_{\mathrm{Dec}}$, with the POD modes of the data, $\boldsymbol{\Phi}^{(\mathrm{phys})}_{\mathrm{True}}$. For brevity, we limit our analysis to the latent space of 18 variables in the turbulent case.
%
We decompose the autoencoder latent  dynamics into proper orthogonal modes, $\boldsymbol{\Phi}^{(\mathrm{lat})}_i$, which we name ``autoencoder principal directions'' to distinguish them from the POD modes of the data. 
Figure~\ref{fig:lat_space_pod}a shows that four principal directions are dominant, as indicated by the change in slope of the energy, and that they contain roughly $97\%$ of the energy of the latent space signal.
We therefore focus our analysis on the four principal directions, from which we obtain the decoded field
\begin{equation}
    \boldsymbol{\hat{q}}_{\mathrm{Dec}4}(t) = \boldsymbol{f}\left(\sum_{i=1}^4 a_i(t) \boldsymbol{\Phi}^{(\mathrm{lat})}_i\right),
\end{equation}
where $a_i(t) = \boldsymbol{\Phi}^{(\mathrm{lat})T}_i\boldsymbol{z}(t)$ and $\boldsymbol{f}(\cdot)$ is the decoder (\S\ref{sec:CNN}). The energy content in the reconstructed flowfield is shown in Figure~\ref{fig:lat_space_pod}b. On the one hand, the full latent space, which consists of eighteen modes, reconstructs accurately the energy content of the first fifty POD modes. On the other hand, $\boldsymbol{\hat{q}}_{\mathrm{Dec}4}$ closely matches the energy content of the first few POD modes of the true flow field, but the error increases for larger numbers of POD modes. This happens because $\boldsymbol{\hat{q}}_{\mathrm{Dec}4}$ contains less information than the true flow field, so that fewer POD modes are necessary to describe it. The results are further corroborated by the scalar product with the physical POD modes of the data (Figure~\ref{fig:modes_pod}). The decoded field obtained from all the principal components accurately describes the  majority of the first fifty physical POD modes, and $\boldsymbol{\hat{q}}_{\mathrm{Dec}4}$ accurately describes the first POD modes. 
These results indicate that only few nonlinear modes contain information about several POD modes.

\begin{figure}
    \centering
    \includegraphics[width=1.\textwidth]{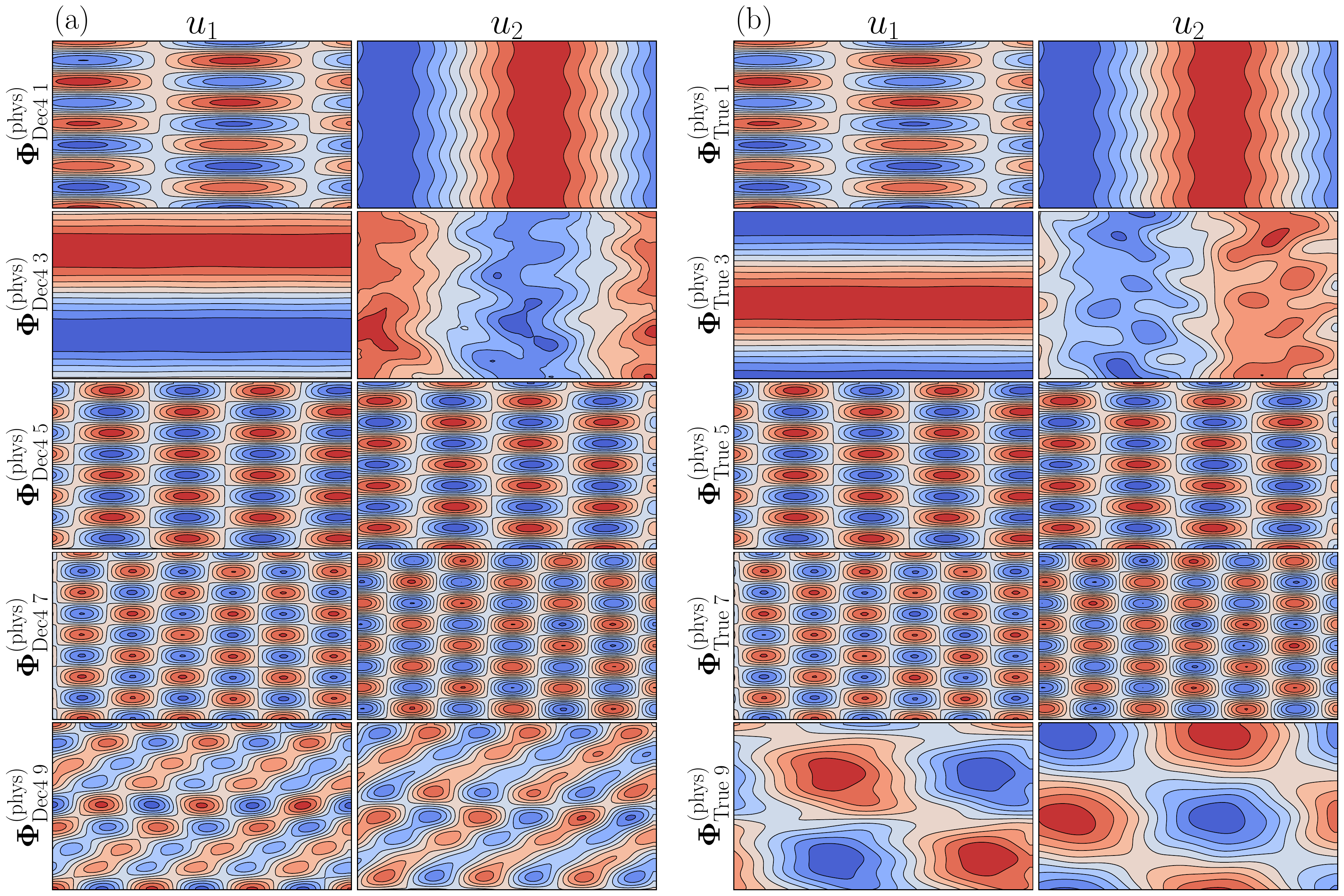}
    \caption{POD modes, one per row, for (a) the reconstructed flow field based on four principal components in latent space and (b) the data. Even modes are shown in the Supplementary material.}
    \label{fig:pods}
\end{figure}

\begin{figure}
    \centering
    \includegraphics[width=1.\textwidth]{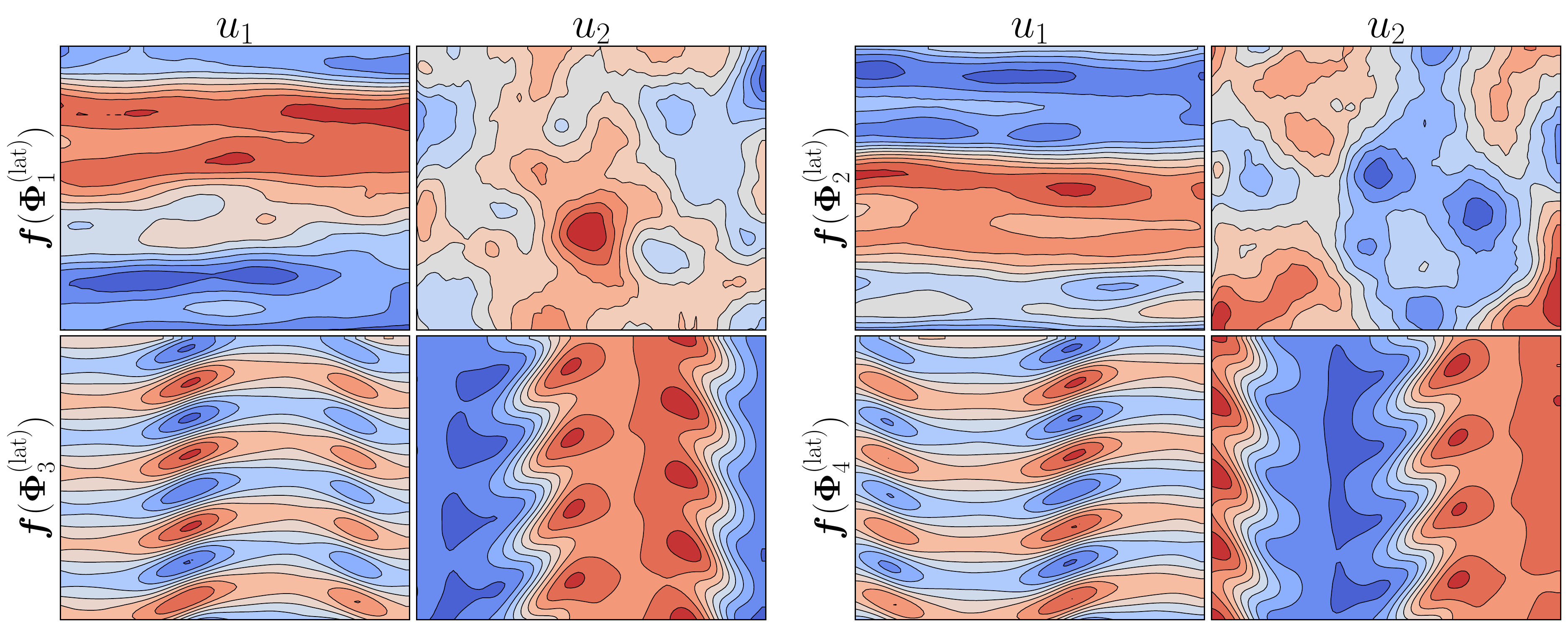}
    \caption{Decoded four principal directions in latent space.}
    \label{fig:decoded_pods}
\end{figure}

A visual comparison of the most energetic POD modes of the true flow field and of $\boldsymbol{\hat{q}}_{\mathrm{Dec}4}$ is shown in Figure~\ref{fig:pods}. The first eight POD modes are accurately recovered.
Each of the decoded latent space principal directions, $\boldsymbol{f}(\boldsymbol{\Phi}^{(\mathrm{lat})}_i)$ is plotted in Figure~\ref{fig:decoded_pods}. We observe that the decoded principal directions are in pairs as the linear POD modes, but they differ significantly from any of the POD modes of Figure~\ref{fig:pods}. This is because the decoder infers nonlinear interactions among the POD modes.

\section{Time-accurate prediction}
\label{sec:time-acc}

\begin{figure}
    \centering
    \includegraphics[width=.5\textwidth]{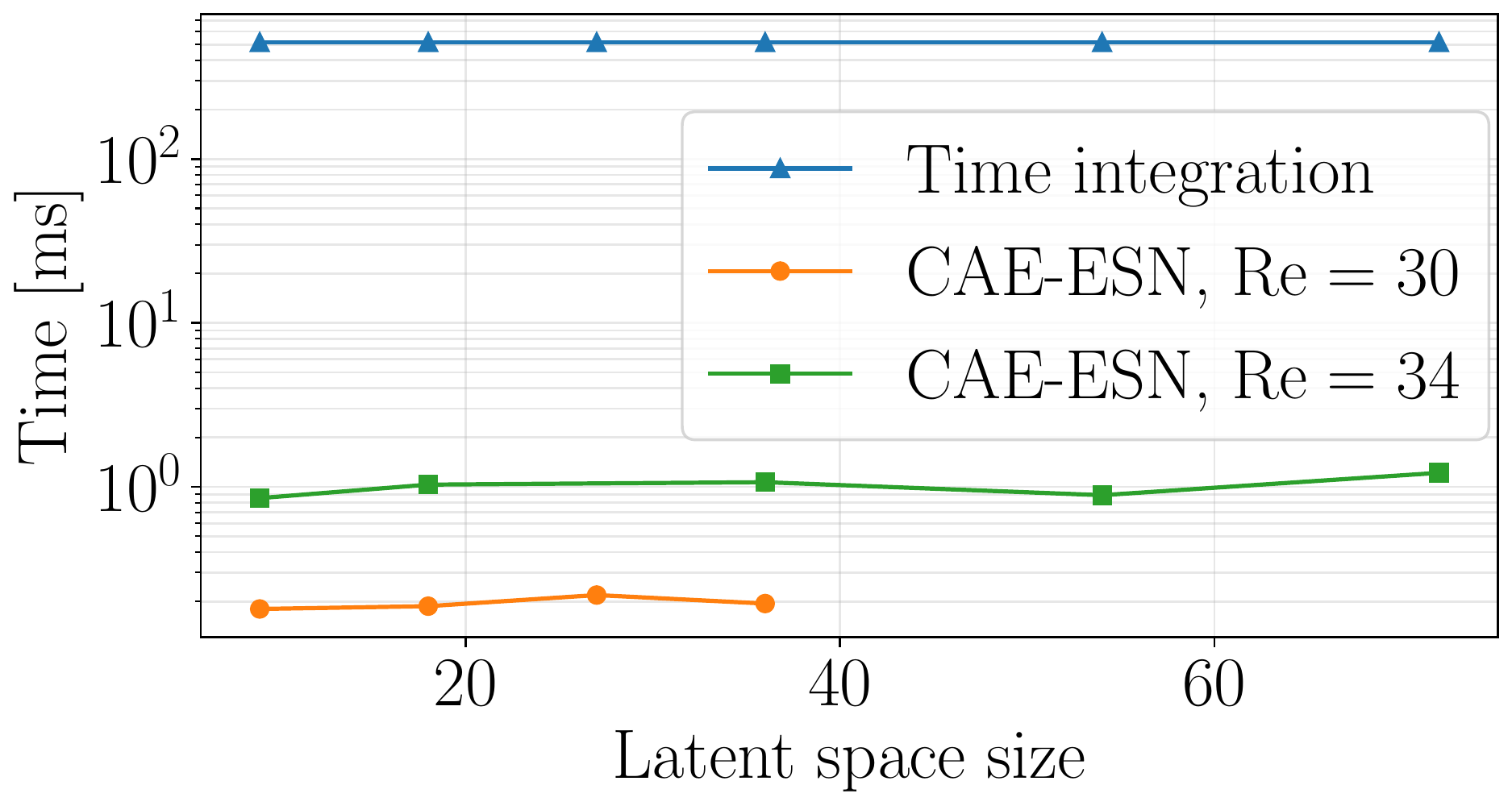}
    \caption{Computational time required to forecast the evolution of the system for 1 time unit. Times for the CAE-ESN are for the largest size of the reservoir, which takes the longest time.
    }
    \label{fig:comp_time}
\end{figure}

Once the autoencoder is trained to provide the latent space, we train an ensemble of ten echo state networks to predict the latent space dynamics. We use an ensemble of networks to take into account the random initialization of the input and state matrices \citep{racca2022data}. We predict the low-dimensional dynamics to reduce the computational cost, which becomes prohibitive when time-accurately predicting high-dimensional systems with recurrent neural networks \citep{vlachas2020backpropagation}. 
Figure~\ref{fig:comp_time} shows the computational time required to forecast the evolution of the system by solving the governing equations and using the CAE-ESN. The governing equations are solved using a single GPU Nvidia Quadro RTX 4000, whereas the CAE-ESN uses a single CPU Intel i7-10700K (ESN) and single GPU Quadro RTX 4000 (decoder) in sequence. The CAE-ESN is two orders of magnitude faster than solving the governing equations because the ESNs use sparse matrix multiplications and can advance in time with a $\delta t$ larger than the numerical solver. 

\subsection{Quasiperiodic case}

\begin{figure}
    \centering
    \includegraphics[width=1.\textwidth]{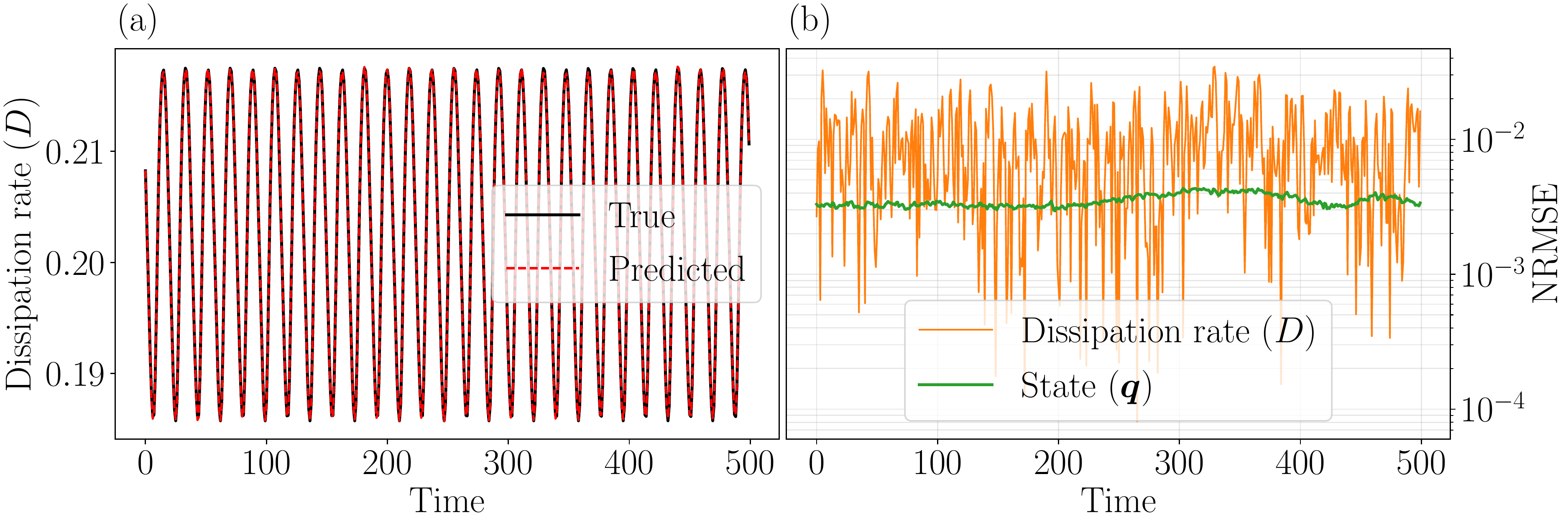}
    \caption{Prediction of the dissipation rate in the quasiperiodic case in the test set, i.e. unseen dynamics, for a CAE-ESN with a 2000 neurons reservoir and nine-dimensional latent space.}
    \label{fig:Dis_Re30_TS}
\end{figure}

\begin{figure}
    \centering
    \includegraphics[width=1.\textwidth]{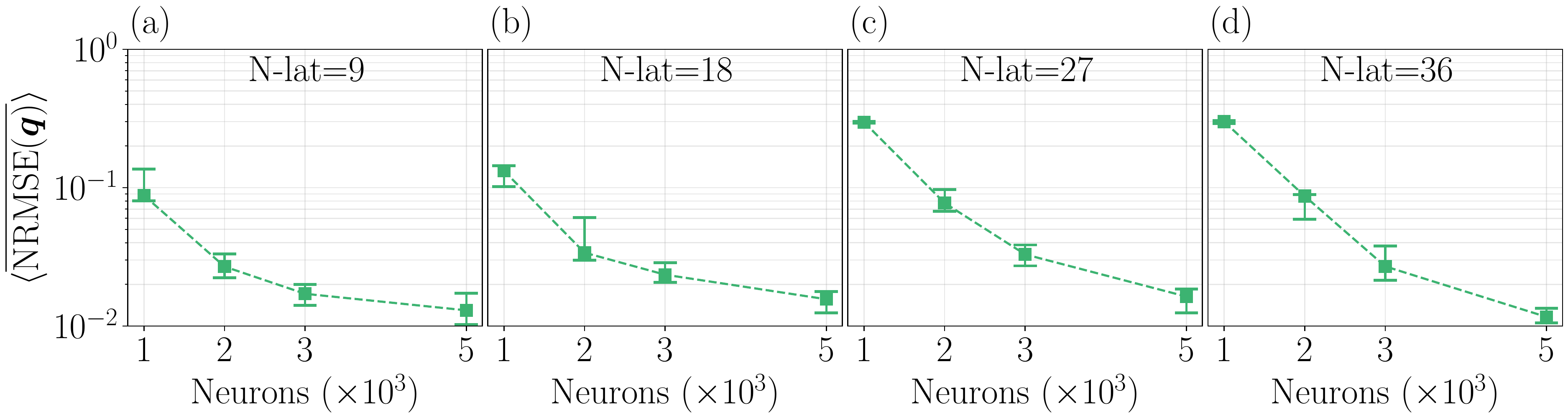}
    \caption{25th, 50th and 75th percentiles of the average NRMSE in the test set as a function of the latent space size and reservoir size in the quasiperiodic case.}
    \label{fig:QP-NRMSE}
\end{figure}

We train the ensemble of echo state networks using 15000 snapshots equispaced by $\delta t = 1$. The networks are validated and tested in closed-loop on intervals lasting 500 time units. One closed-loop prediction of the average dissipation rate \eqref{eq:dissipation} in the test set is plotted in Figure~\ref{fig:Dis_Re30_TS}. The network accurately predicts the dissipation rate (Figure~\ref{fig:Dis_Re30_TS}a) for several periods. The NRMSE of the dissipation and the state vector oscillates around values smaller than $10^{-2}$ throughout the prediction (Figure~\ref{fig:Dis_Re30_TS}b).
Figure~\ref{fig:QP-NRMSE} shows the quantitative results for different latent spaces and reservoir sizes. We plot the percentiles of the network ensemble for the mean over 50 intervals in the test set, $\langle\cdot\rangle$, of the time-averaged NRMSE. The CAE-ESN time-accurately predicts the system in all cases analysed, except for small reservoirs in large latent spaces (Figure \ref{fig:QP-NRMSE}c,d). This is because larger reservoirs are needed to accurately learn the dynamics of larger latent spaces. For all the different latent spaces, the accuracy of the prediction increases with the size of the reservoir. For 5000 neurons reservoirs, the NRMSE for the prediction in time is the same order of magnitude as the NRMSE of the reconstruction (Figure~\ref{fig:Rec}). This means that the CAE-ESN learns the quasiperiodic dynamics and accurately predicts the future evolution of the system for several characteristic timescales of the system.


\subsection{Turbulent case}

\begin{figure}
    \centering
    \includegraphics[width=1.\textwidth]{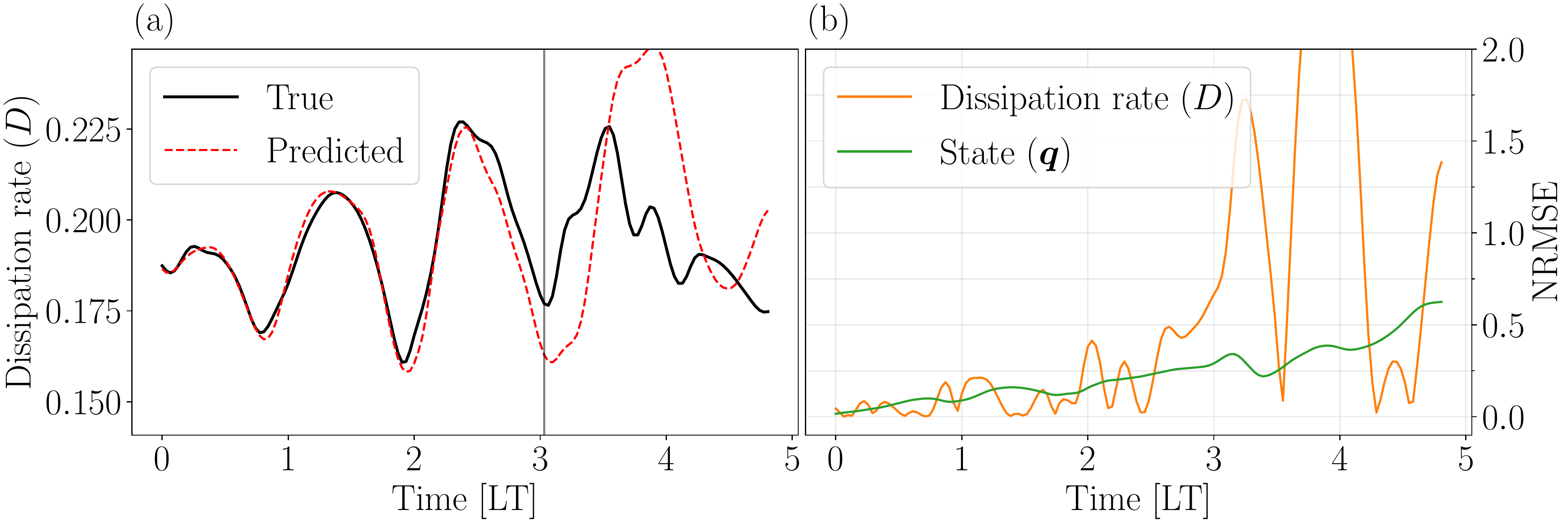}
    \caption{
    Prediction of the dissipation rate in the turbulent case in the test set, i.e. unseen dynamics, for a CAE-ESN with a 10000 neurons reservoir and 36-dimensional latent space. The vertical line in panel (a) indicates the prediction horizon.}
    \label{fig:Dis_Re34_TS}
\end{figure}

\begin{figure}
    \centering
    \includegraphics[width=.8\textwidth]{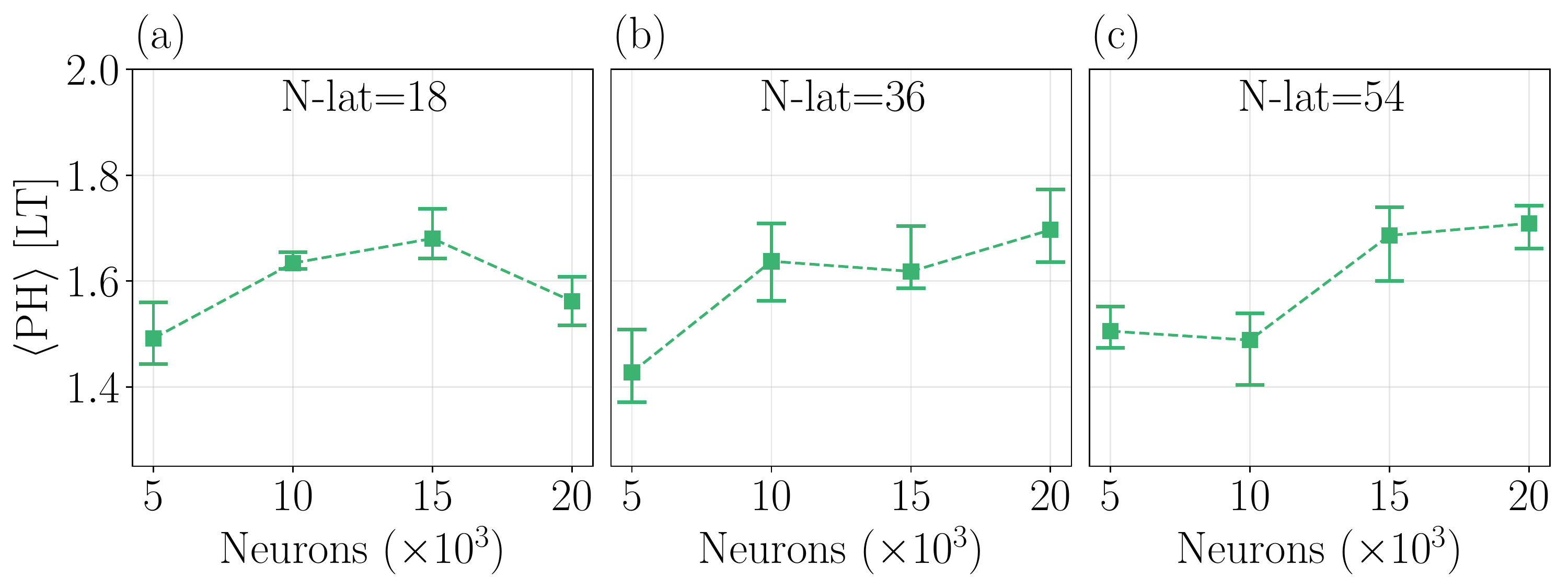}
    \caption{25th, 50th and 75th percentiles of the average prediction horizon in the test set as a function of the latent space and reservoir size in the turbulent case. 
    }
    \label{fig:CH-PH}
\end{figure}

We validate and test the networks for the time-accurate prediction of the turbulent dynamics against the Prediction Horizon (PH). The prediction horizon is the time interval during which the NRMSE \eqref{eq:NRMSE} is smaller than a user-defined threshold, $k$
\begin{equation}
    \mathrm{PH} = \mathop{\mathrm{argmax}}_{t_p}(t_p \,|\, \mathrm{NRMSE}(\boldsymbol{q}) < k),
\end{equation}
where $t_p$ is the time from the start of the closed-loop prediction and $k=0.3$. The prediction horizon is a commonly used metric, which is tailored for the prediction of diverging trajectories in chaotic (turbulent) dynamics \citep{boffetta2002predictability, vlachas2020backpropagation}. The prediction horizon is evaluated in Lyapunov times, LTs (\S\ref{sec:data}).
Figure~\ref{fig:Dis_Re34_TS}a shows the prediction of the dissipation rate in an interval of the test set. The predicted trajectory closely matches the true data for about three Lyapunov Times. Owing to the intrinsic nature of chaos, as expected, the prediction error ultimately increases with time (Figure~\ref{fig:Dis_Re34_TS}b). 
%
To quantitatively assess the performance of the CAE-ESN, we create twenty latent spaces for each of the latent space sizes $[18,36,54]$. Each latent space is obtained by training  autoencoders with different random initialisations and optimisations of the weights of the convolutional layers (\S\ref{sec:CNN}). The size of the latent space is selected to be larger than the Kaplan-Yorke dimension of the system, $N_{KY} = 9.5$ (\S\ref{sec:KY}) in order to contain the attractor's dynamics. In each latent space, we train the ensemble of ten echo state networks using 60000 snapshots equispaced by $\delta t =0.5$.
Figure~\ref{fig:CH-PH} shows the average prediction horizon over 100 intervals in the test set for the best performing latent space of each  size.
The CAE-ESN successfully predicts the system for up to more 1.5 LTs for all sizes of the latent space. Increasing the reservoir size slightly improves the performance with a fixed latent space size, whilst latent spaces of different sizes perform similarly. This means that small reservoirs in small latent spaces are sufficient to time-accurately predict the system. 
%
%
%

\begin{figure}
    \centering
    \includegraphics[width=1.\textwidth]{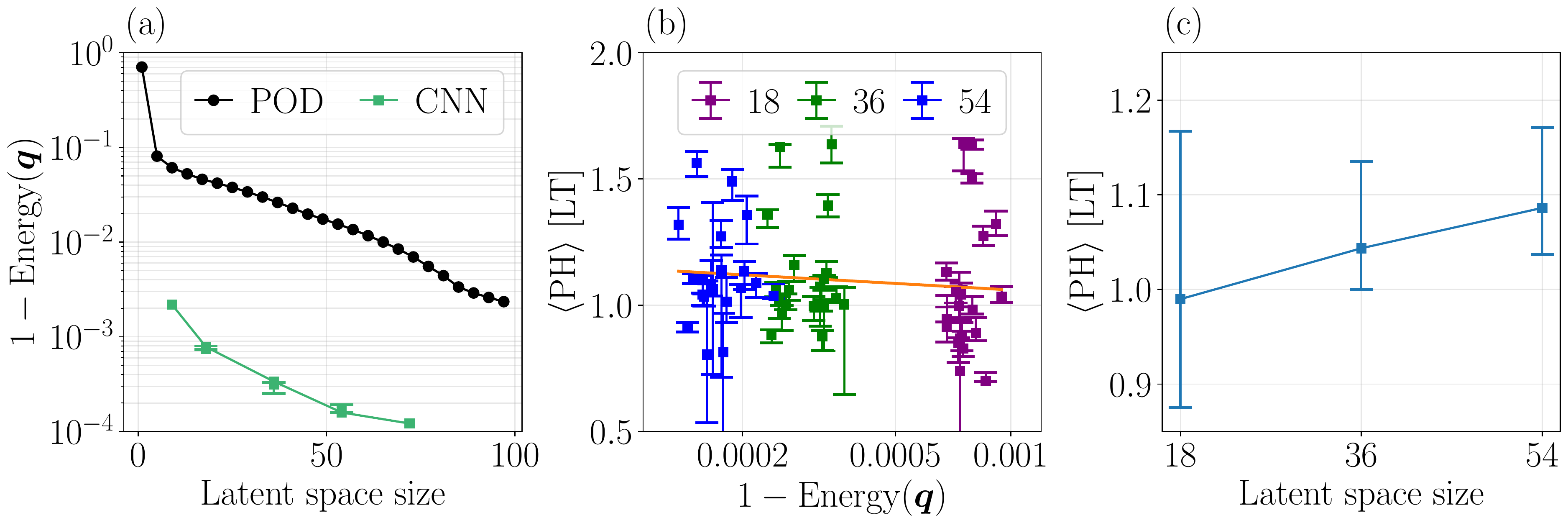}
    \caption{25th, 50th and 75th percentiles of (a) the reconstruction error of Figure \ref{fig:Rec} from twenty different latent spaces of different size, (b) the PH as a function of the reconstruction error, and (c) the medians of the PH of (b) as a function of the latent space size. The line in (b) is obtained by linear regression.}
    \label{fig:latent_spaces_timeacc}
\end{figure}

A detailed analysis of the performance of the ensemble of the latent spaces of the same size is shown in Figure~\ref{fig:latent_spaces_timeacc}.  Figure~\ref{fig:latent_spaces_timeacc}a shows the reconstruction error. The small amplitude of the errorbars indicates that different autoencoders with same latent space size reconstruct the flowfield with similar energy. Figure \ref{fig:latent_spaces_timeacc}b shows the performance of the CAE-ESN with 10000 neurons reservoirs for the time-accurate prediction  as a function of the reconstruction error. In contrast to the error in Figure~\ref{fig:latent_spaces_timeacc}a, the prediction horizon varies significantly, ranging from 0.75 to more than 1.5 Lyapunov Times in all latent space sizes. 
Moreover, there is no discernible correlation between the prediction horizon and reconstruction error.
This means that the time-accurate performance depends more on the training of the autoencoder than the latent space size.
To quantitatively assess the correlation between two quantities, $[\boldsymbol{a_1},\boldsymbol{a_2}]$,
we use the Pearson correlation coefficient \citep{galton1886regression,pearson1895vii}
\begin{equation}
     r(\boldsymbol{a_1},\boldsymbol{a_2}) = \frac{\sum_i(a_{1_i}-\langle a_1 \rangle)(a_{2_i}-\langle a_2 \rangle)}
     {\sqrt{\sum_i(a_{1_i}-\langle a_1 \rangle)^2}\sqrt{\sum_i(a_{2_i}- \langle a_2 \rangle)^2}}.
\end{equation}
The values $r=\{-1,0,1\}$ indicate anticorrelation, no correlation and correlation, respectively. The Pearson coefficient between the reconstruction error and the prediction horizon is $r=-0.10$ for the medians of the sixty cases analysed in Figure~\ref{fig:latent_spaces_timeacc}b, which means that the prediction horizon is weakly correlated with the reconstruction error. This indicates that latent spaces that similarly capture the energy of the system, may  differ in capturing the dynamical content of the system; i.e., some modes that are critical for the dynamical evolution of the system do not necessarily affect the reconstruction error, and therefore may not necessarily be captured by the autoencoder. This is a phenomenon that also affects different reduced-order modelling methods,
specifically POD \citep{Rowley2017, agostini2020exploration}.
In Figure~\ref{fig:latent_spaces_timeacc}c, we plot the prediction horizon of the medians of the errorbars of Figure~\ref{fig:latent_spaces_timeacc}b for different latent spaces. 
Although the performance varies within the same size of the latent space, we observe that the prediction horizon improves and the range of the errorbars decreases with increasing size of the latent space. This is because autoencoders that better approximate the system in a $L_2$ sense are also  more likely to include relevant dynamical information for the time evolution of the system. 
From a design perspective, this means that latent spaces with smaller reconstruction error are needed to improve and reduce the variability of the prediction of the system. A different approach, which is beyond the scope of this paper, is to generate latent spaces and metrics tailored for time-accurate prediction (see Supplementary material).

\section{Statistical prediction}
\label{sec:statistics}

We analyse the statistics predicted by the CAE-ESN with long-term predictions. Long-term predictions are closed-loop predictions that last tens of Lyapunov times, whose instantaneous state differs from the true data because of chaos (infinitesimal errors exponentially grow in chaotic systems). However, in ergodic systems, the trajectories evolve within a bounded region of the phase space, the attractor, which means that they share the same long-term statistics.
The long-term predictions are generated from 50 different starting snapshots in the training set. In the quasiperiodic case, each time series is obtained by letting the CAE-ESN evolve in closed-loop for 2500 time units, whereas in the chaotic case the CAE-ESN evolves for 30 LTs. 
(If the network predicts values outside the range of the training set, we discard the remaining part of the closed-loop prediction, similarly to~\citet{racca2022data}.)
To quantify the error of the predicted PDF with respect to the true PDF, we use the first-order Kantorovich Metric \citep{kantorovich1942translocation}, also known as the Wasserstein distance or Earth mover's distance. The Kantorovich metric evaluates the work to transform one PDF to another, which is computed as
\begin{equation}
    \mathcal{K} = \int_{-\infty}^{\infty}|\mathrm{CDF}_1(k) - \mathrm{CDF}_2(k)|dk,
\end{equation}
where $\mathrm{CDF}$ is the Cumulative Distribution Function and $k$ is the random variable of interest. Small values of $\mathcal{K}$ indicate that the two PDFs are similar to each other, whereas large values of $\mathcal{K}$ indicate that the two PDFs significantly differ from each other.

\subsection{Quasiperiodic case}

\begin{figure}
    \centering
    \includegraphics[width=1.\textwidth]{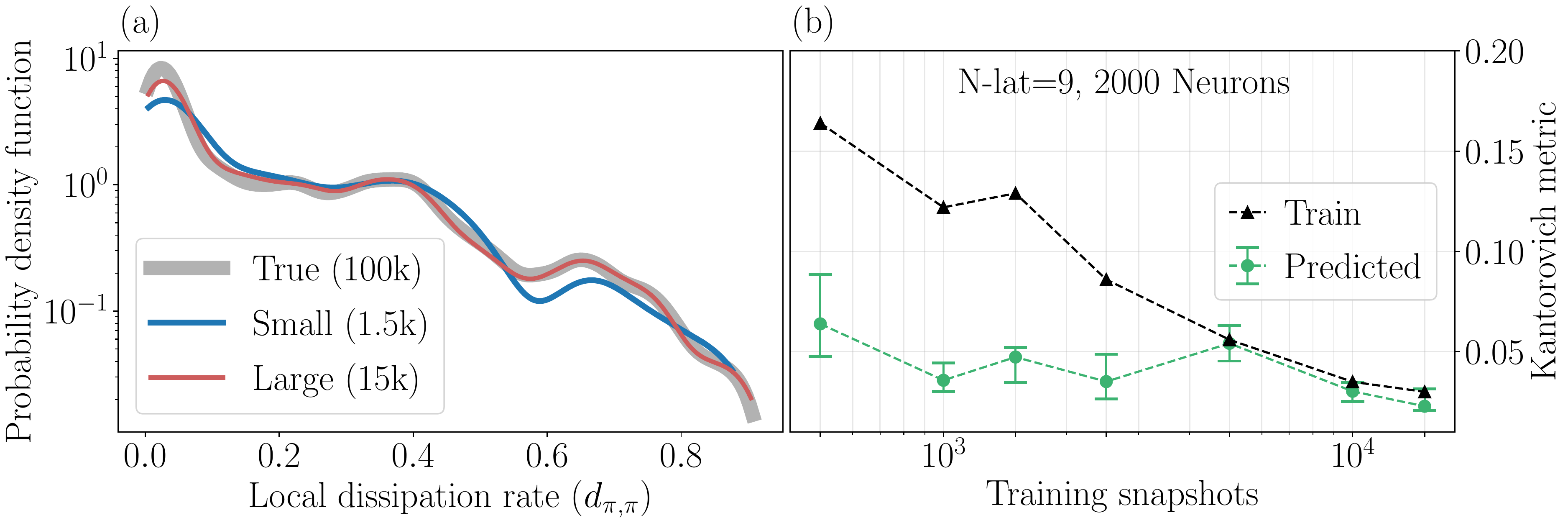}
    \caption{(a) Probability density function of the local dissipation rate, $d_{\pi,\pi}$, in the training set for different numbers of training snapshots. (b) 25th, 50th and 75th percentiles of the Kantorovich metric for the training set (Train) and for the CAE-ESN (Predicted) as a function of the training set size. Quasiperiodic regime.
    }
    \label{fig:Dis-Stats-30-Trains}
\end{figure}

\begin{figure}
    \centering
    \includegraphics[width=1.\textwidth]{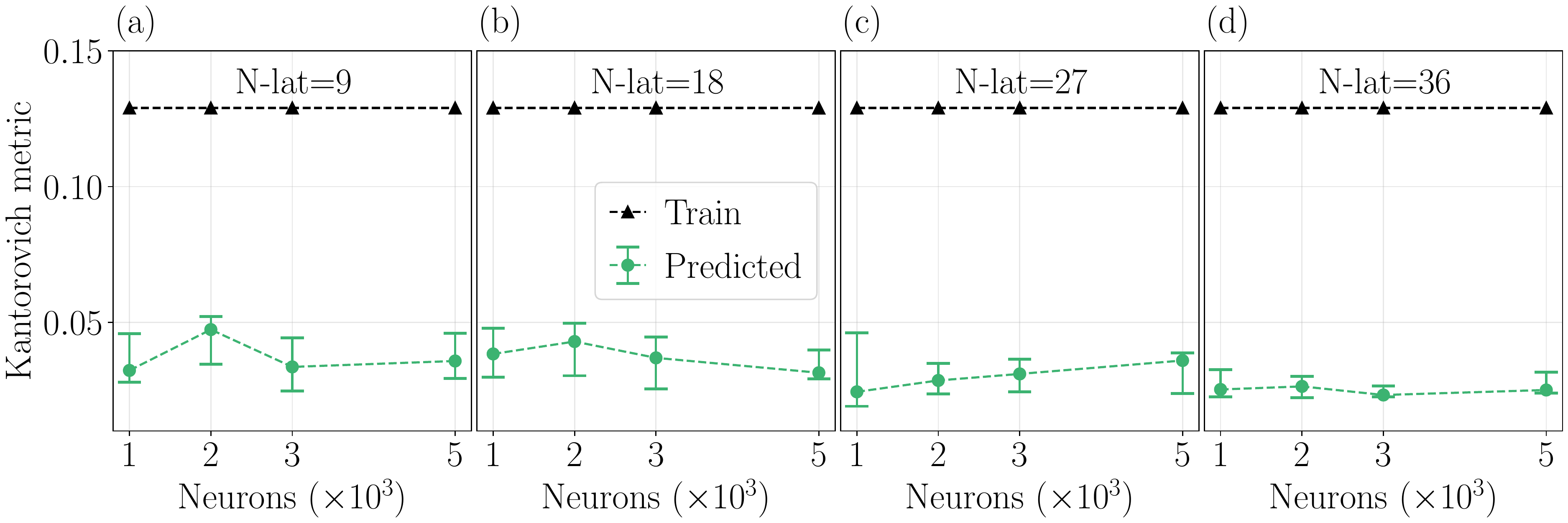}
    \caption{25th, 50th and 75th percentiles of the Kantorovich metric of the local dissipation rate, $d_{\pi,\pi}$, for the training set (Train) and for the networks (Predicted) as a function of the latent space and reservoir size in the quasiperiodic case. 
    }
    \label{fig:Dis-Stats-30}
\end{figure}

To predict the statistics of the quasiperiodic case, we train the network on small datasets. We do so because the statistics are converged for the large dataset (15000 time units) used for training in \S~\ref{sec:time-acc} (Figure~\ref{fig:Dis-Stats-30-Trains}a).
We use datasets with non-converged statistics because here our objective is to increase the statistical knowledge of the system with respect to the imperfect data available during training. We analyse the local dissipation rate at the central point of the domain, $d_{\pi,\pi}$, because the average dissipation rate is periodic (\S~\ref{sec:data}). The statistics of the periodic signal converge fast, whereas the statistics of the quasiperiodic components of the system's state need larger intervals.
Figure~\ref{fig:Dis-Stats-30-Trains}b shows the Kantorovich metric for the CAE-ESN as a function of the number of training snapshots in the dataset for fixed latent space and reservoir size.
The networks improve the statistical knowledge of the system with respect to the training set in all cases analysed. The improvement is larger in smaller datasets, and it saturates for larger datasets. This means that (i) the networks accurately learn the statistics from small training sets, and  (ii) the networks are able to extrapolate from imperfect statistical knowledge and provide an accurate estimation of the true statistics. Figure~\ref{fig:Dis-Stats-30} shows the performance of the CAE-ESN as a function of the reservoir size for the small dataset of 1500 training snapshots. In all cases analysed, the CAE-ESN outperforms the training set, with values of the Kantorovich metric up to three times smaller than those of the training set. Consistently with previous studies on ESNs \citep{huhn2021gradient, racca2022data}, the performance is approximately constant for different reservoir sizes. The results indicate that small reservoirs and small latent spaces are sufficient to accurately extrapolate the statistics of the flow from imperfect datasets. 

\subsection{Turbulent case}
\begin{figure}
    \centering
    \includegraphics[width=1.\textwidth]{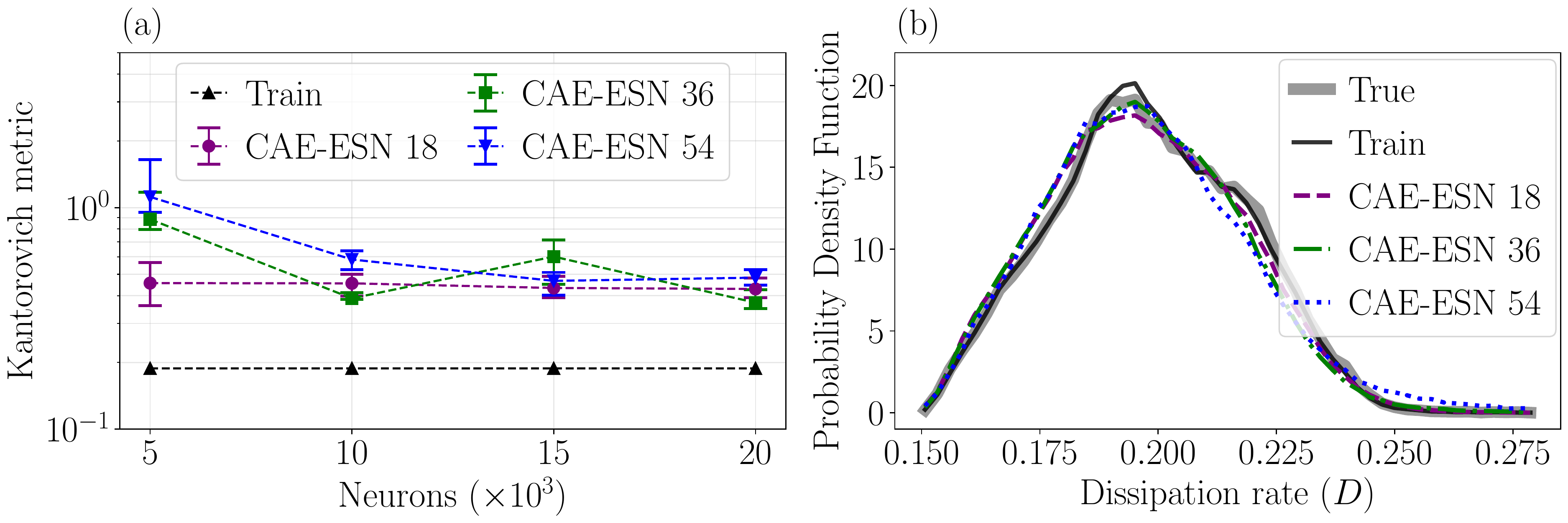}
    \caption{(a) 25th, 50th and 75th percentiles of the Kantorovich metric of the  dissipation rate, for the training set (Train), and for different reservoir and latent space sizes. (b) PDF for the true and training data, and predicted by 10000 neurons CAE-ESNs. 
    }
    \label{fig:Ch-Stats}
\end{figure}

In the turbulent case, we analyse the ability of the architecture to replicate the PDF of the average dissipation rate, $D$, to assess whether the networks have learned the dynamics in a statistical sense. 
Figure~\ref{fig:Ch-Stats} shows the results for the best performing CAE-ESN with latent space of size 18, 36 and 54 discussed in \S~\ref{sec:time-acc}. Figure~\ref{fig:Ch-Stats}a shows the Kantorovich metric as a function of the size of the reservoir. The networks predict the statistics of the dissipation rate with Kantorovich metrics of the same order of magnitude of the training set, which has nearly converged statistics. This means that the network accurately predict the PDF of the dissipation rate  (Figure~\ref{fig:Ch-Stats}b). In agreement with the quasiperiodic case, the performance of the networks is approximately constant as a function of the reservoir size. These results indicate that small latent spaces and small networks can learn the statistics of the system.
%
%

\begin{figure}
    \centering
    \includegraphics[width=1.\textwidth]{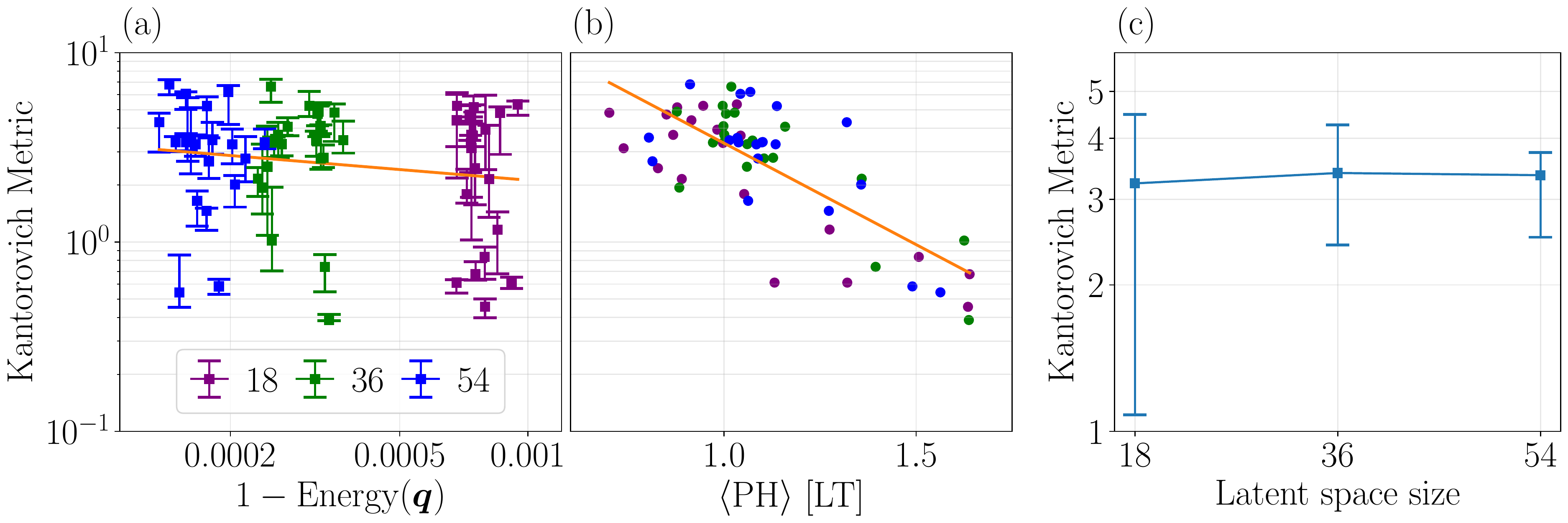}
    \caption{(a) 25th, 50th and 75th percentiles for the Kantorovich metric as a function of the reconstruction error, (b) medians of the Kantorovich metric as a function of the median of the prediction horizon, and (c) 25th, 50th and 75th percentiles of the medians of the Kantorovich metric of (b) as a function of the latent space. The lines in (a)-(b) are obtained by linear regression.}
    \label{fig:latent_spaces_stats}
\end{figure}

A detailed analysis of the statistical performance of the 20 latent spaces of size $[18,36,54]$ is shown in Figure~\ref{fig:latent_spaces_stats}. On the one hand, Figure~\ref{fig:latent_spaces_stats}a shows the scatter plot between the Kantorovich metric and the reconstruction error. In agreement with the results for the prediction horizon (see Figure~\ref{fig:latent_spaces_timeacc}), we observe that (i) the Kantorovich metric varies significantly within latent spaces of the same size, (ii) the two quantities are weakly-correlated ($r=-0.16$) and (iii) the variability of the performance decreases with increasing latent space size (Figure~\ref{fig:latent_spaces_stats}c). On the other hand, Figure \ref{fig:latent_spaces_stats}b shows the scatter plot for the Kantorovich metric as a function of the prediction horizon, which represents the correlation between the time-accurate and statistical performances of the networks. The two quantities are (anti)correlated, with a Pearson coefficient $r=-0.76$.
Physically, the results indicate that (i)  increasing the size of the latent space is beneficial for predicting the statistics of the system, and (ii)  the time-accurate and statistical performances of latent spaces are correlated. This means that the design guidelines for the time-accurate prediction discussed in \S \ref{sec:time-acc} extend to the design of latent spaces for the statistical prediction of turbulent flowfields. 


\section{Conclusions}
\label{sec:conc}

We propose the convolutional autoencoder echo state network (CAE-ESN) for the  spatiotemporal decomposition and prediction of turbulent flowfields from data. 
The convolutional autoencoder nonlinearly maps the  flowfields onto (and back from) the latent space, whose dynamics are predicted by an echo state network.
We deploy the CAE-ESN to analyse a two dimensional  turbulent flow, in both quasiperiodic and turbulent regimes. 
First, we show that the CAE-ESN time-accurately and statistically predicts the flow. 
The latent space requires less than 1\% of the degrees of freedom of the original flowfield.
In the prediction of the spatiotemporal dynamics, the CAE-ESN is at least a hundred time faster than solving the governing equations. This is possible thanks to the nonlinear compression provided by the autoencoder, which has a reconstruction error that is less than 1\% of that from Proper Orthogonal Decomposition.
Second, we analyse the performance of the architecture at different Reynolds numbers. The architecture correctly learns the dynamics of the system in both quasiperiodic and turbulent testcases. This means that the architecture is robust with respect to changes in the physical parameters of the system.
Third, we analyse the performance of the CAE-ESN as a function of the reconstruction error. We show that (i) the performance varies  between latent spaces with similar reconstruction error, (ii) larger latent spaces with smaller reconstruction errors provide a more consistent and more accurate prediction on average, and (iii) the time-accurate and statistical performance of the CAE-ESN are correlated. This means that relatively small reservoirs in relatively small latent spaces are sufficient to time-accurately and statistically predict the turbulent dynamics, and increasing the latent space size is beneficial for the performance of the CAE-ESN. 
By proposing a data-driven framework for the nonlinear decomposition and prediction of  turbulent spatiotemporal dynamics, this paper opens up possibilities for nonlinear reduced-order modelling of turbulent flows.


\section*{Acknowledgements}

A.Racca is supported by the Engineering and Physical Sciences Research Council under the Research Studentship No 2275537 and by the Cambridge Commonwealth, European \& International Trust under a Cambridge European Scholarship.
%
%
L. Magri gratefully acknowledges financial support from the ERC Starting Grant PhyCo 949388. The authors are grateful to Daniel Kelshaw for helpful discussions on the numerical solver.  

\section*{Declaration of interests } The authors report no conflict of interest.

\appendix

\section{Autoencoder layers}

\label{sec:tab_layers}

We provide additional details about the autoencoder described in \S \ref{sec:CNN}, Table \ref{tab:CNN_layers}. 
The total number of trainable parameters of the autoencoder is around 200k (the exact number varies depending on the size of the latent space).
\begin{table}
\centering
\begin{tabular}{l l l}
Layer Type $\qquad$ & Notes  & Output size \\
\midrule
\multicolumn{3}{c}{Encoder} \\
\midrule
Periodic Padding & Each scale has different padding $\qquad$ & $(49,49,2)$ \\
Convolution & Stride=2  & $(24,24,6)$ \\
Periodic Padding & Each scale has different padding $\qquad$ & $(26,26,6)$ \\
Convolution & Stride=1 $\qquad$ & $(24,24,6)$ \\
Periodic Padding & Each scale has different padding $\qquad$ & $(25,25,6)$ \\
Convolution & Stride=2 & $(12,12,12)$ \\
Periodic Padding & Each scale has different padding $\qquad$ & $(14,14,12)$ \\
Convolution &  Stride=1 $\qquad$ & $(12,12,12)$ \\
Periodic Padding & Each scale has different padding $\qquad$ & $(13,13,12)$ \\
Convolution      & Stride=2 & $(6,6,24)$ \\
Periodic Padding & Each scale has different padding $\qquad$ & $(8,8,24)$ \\
Convolution & Stride=1 $\qquad$ & $(6,6,24)$ \\
Periodic Padding & Each scale has different padding $\qquad$ & $(7,7,24)$ \\
Convolution & Stride=2 and varying kernel depth & $(3,3,\frac{N_{\mathrm{lat}}}{9})$ \\
\midrule
\multicolumn{3}{c}{Decoder} \\
\midrule
Periodic Padding & All scales have equal padding  & $(5,5,\frac{N_{\mathrm{lat}}}{9})$ \\
Transpose Convolution & Stride=2, output size varies with scale & $(11,11,24)$ \\
Convolution & Stride=1 & $(9,9,24)$ \\
Transpose Convolution & Stride=2, output size varies with scale & $(19,19,12)$ \\
Convolution & Stride=1 & $(17,17,12)$ \\
Transpose Convolution & Stride=2, output size varies with scale & $(35,35,6)$ \\
Convolution & Stride=1 & $(33,33,6)$ \\
Transpose Convolution & Stride=2, output size varies with scale & $(67,67,3)$ \\
Center Crop & Cropped to have right padding & $(50,50,3)$ \\
Convolution & Stride=1, linear activation & $(48,48,2)$
\\
\end{tabular}
\caption{Autoencoder layers. Output size differs from one scale to the other, here shown the ones relative to the $3\times3$ filter.}
\label{tab:CNN_layers}
\end{table}
\section{Echo state networks hyperparameters}
\label{app:esn_val}

We provide additional details regarding the selection of hyperparameters in echo state networks.
In the quasiperiodic case, we explore the hyperparameter space $[0.01,5]\times[0.8,1.2]$ for $[\sigma_{\mathrm{in}},\rho]$ in logarithmic scale for the input scaling. We optimize the spectral radius in the range around unity from previous studies with quasiperiodic data \citep{racca2021robust,novoa2021real}. We select the hyperparameters to minimize the average MSE of 30 intervals of length 500 time units taken from the training and validation dataset. To select the noise to be added to the inputs, for each network we perform a search in $k_z$ in points equispaced in logarithmic scale by $\log_{10}(\sqrt{10})$. The search starts by evaluating the average MSE in the test set for $[k_{z1},k_{z2}] = [10^{-3}, \sqrt{10}\times10^{-3}]$. Based on the slope of the MSE as a function of $k_z$ provided by $[k_{z1},k_{z2}]$, we select $k_{z3}$ to either be $k_{z1}/\sqrt{10}$ or $k_{z2}\times\sqrt{10}$ in order to minimize the MSE. From $[k_{z1},k_{z2}, k_{z3}]$ we select a new point, and so on and forth. The search continues until a local minimum of the MSE is found or the maximum numbers of four function evaluations is reached.
In the chaotic case, we explore the hyperparameter space $[0.1,5]\times[0.1,1.]$ for $[\sigma_{\mathrm{in}},\rho]$ in logarithmic scale for the input scaling. We select the hyperparameters to maximize the average prediction horizon of 50 starting points taken from the training and validation dataset. To select the noise to be added to the inputs, we follow a similar procedure to the quasiperiodic case. However, due to computational constraints related to larger sizes of the reservoir, we perform the search only for the first network of the ensemble and utilize the optimal $k_z$ for all the other networks in the ensemble. The search starts from $[k_{z1},k_{z2}] = [\sqrt{10}\times10^{-3}, 10^{-2}]$.
In both cases, (i) washout consists of fifty steps, (ii) we optimize the grid $[10^{-3}, 10^{-6},10^{-9}]$ for $\beta$, and (iii) the Bayesian optimization starts from a grid of $5\times5$ points in the $[\sigma_{\mathrm{in}},\rho]$ domain and then selects five additional points through the expected improvement acquisition function \citep{brochu2010tutorial}.
The Bayesian optimization is implemented using Scikit-learn \citep{scikit-learn}.
 
\begin{figure}
    \centering
    \includegraphics[width=.5\textwidth]{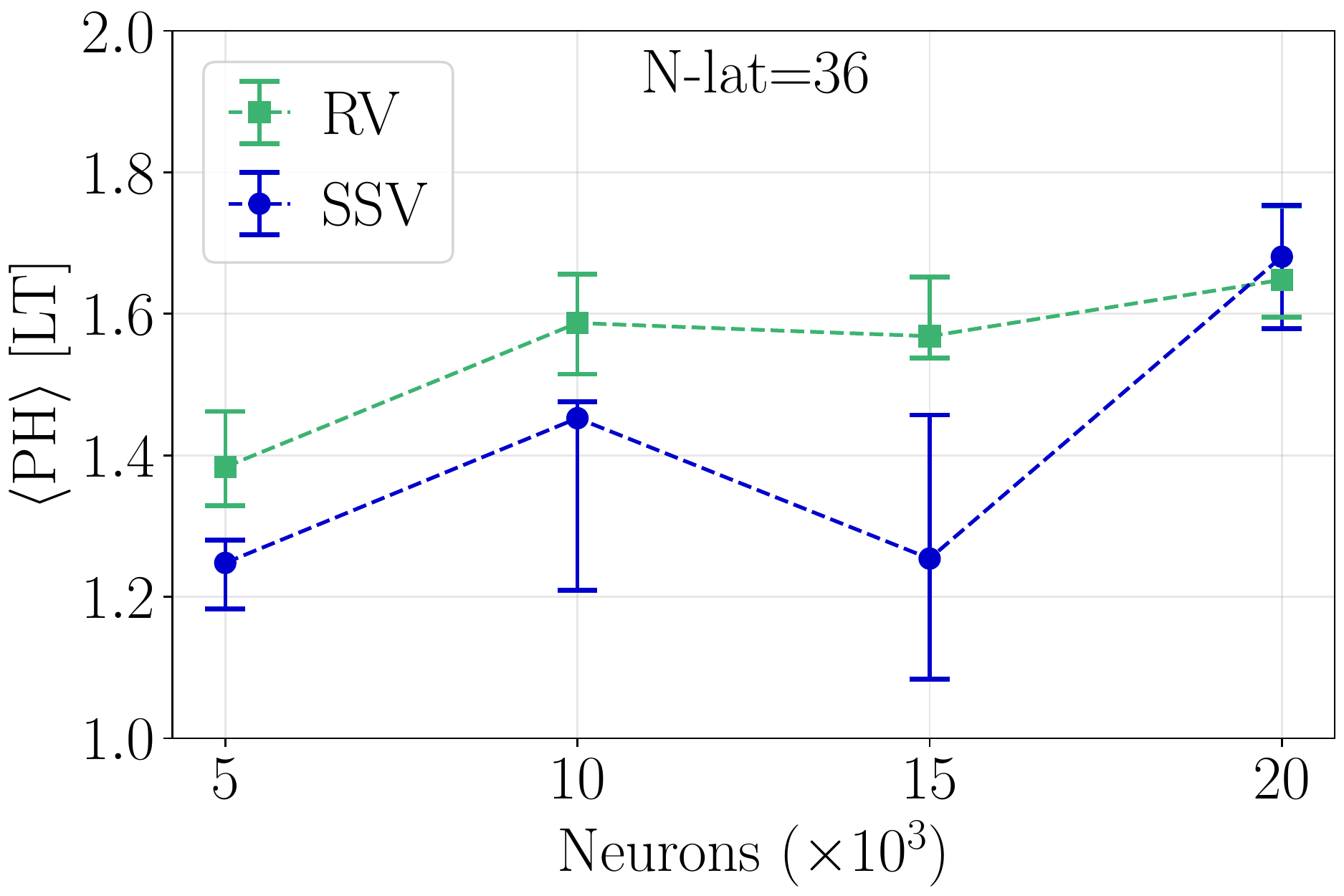}
    \caption{25th, 50th and 75th percentiles of the average prediction horizon in the test set for the recycle validation (RV) and the Single Shot Validation (SSV). 
    }
    \label{fig:CH-PH-SSv}
\end{figure}

We show a comparison of the recycle validation with the single shot validation in Figure~\ref{fig:CH-PH-SSv} for the prediction of the turbulent case. The recycle validation outperforms the single shot validation, providing an increase of roughly 10\% on average of the 50th percentile of the Prediction Horizon. Moreover, the recycle validation reduces the variability of the performance of different networks of the same size, making the training more robust to the random initialization of the networks. This results in the average uncertainty given by the width of the error bars is roughly 70\% larger in the SSV with respect to the RV. These results indicate that the recycle validation outperforms and is more reliable than the single shot validation in the prediction of the turbulent  flow.

\bibliographystyle{jfm}

\clearpage
\setcounter{page}{1}
\setcounter{section}{18}
\renewcommand{\thesection}{\Alph{section}}%
\section{Supplementary material}

\subsection{Selection of number of Fourier modes}
\label{app:spat_convergence}
\begin{figure}
    \centering
    \includegraphics[width=1.\textwidth]{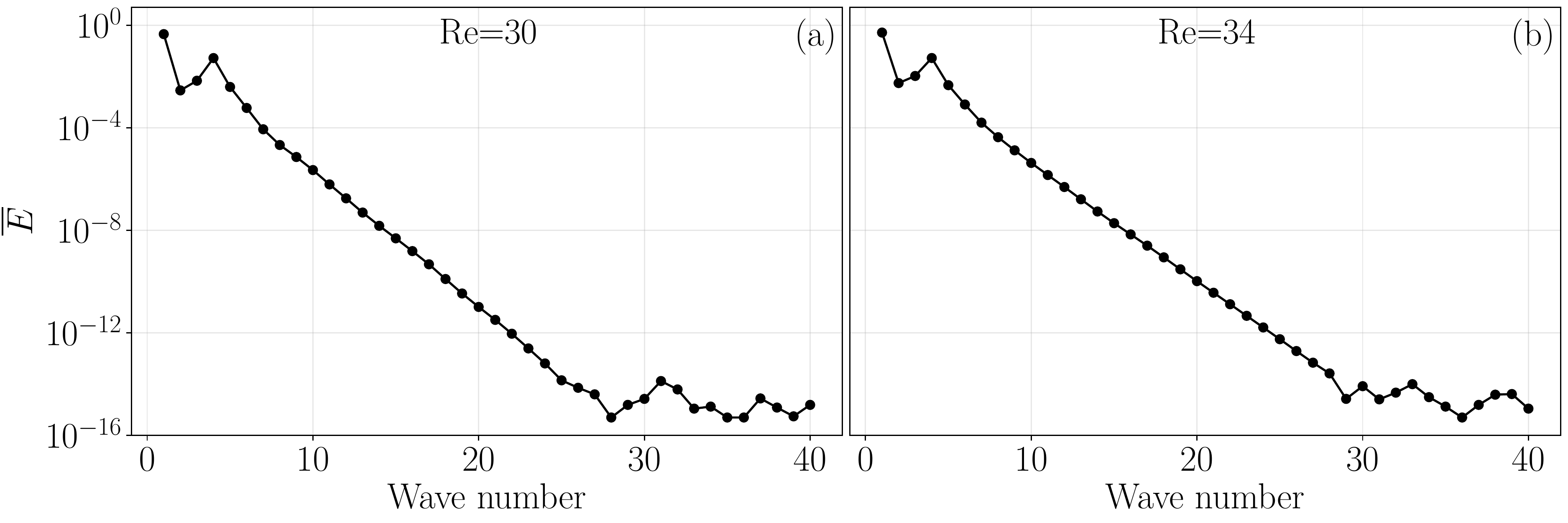}
    \caption{Energy spectra at Re=30, (a), and Re=34, (b).}
    \label{fig:E-k}
\end{figure}

Following \citet{farazmand_2016_adjoint}, we select the number of Fourier modes by studying the convergence of the time-average, $\overline{(\cdot)}$, of the kinetic energy, 
\begin{equation}
    E = \frac{1}{(2\pi)^2} \int_0^{2\pi} \int_0^{2\pi} \frac{1}{2}|| \boldsymbol{u} ||^2 \; \mathrm{d}x_1\mathrm{d}x_2,
\end{equation}
as a function of number of the fourier modes in the numerical solver. 
In both the quasiperiodic and chaotic cases, we select 32 wave numbers, Figure \ref{fig:E-k}.
For $k=32$ wave numbers,
we solve for $(2k +1)^2$ modes, $[-k,-k+1,\dots,0,\dots,k-1,k$] in each direction, where each mode consists of two complex components. Because the physical flowfield has real values, the coefficients of the negative wave numbers are the conjugates of the coefficient of the positive wave numbers, $c_{k1,k2} = c_{-k1,-k2}^*$. This results in $(2k+2)\times k \times 2 \times 2 = 8448$ active degrees of freedom of the system ($c_{0,0}$ is constant as it represents the spatial mean of the flow field) \citep{pope2000turbulent}.

\subsection{Jacobian-free computation of the Lyapunov exponents}

In this section, we describe the algorithm used to compute the first $m$ largest Lyapunov exponents of the system. The algorithm requires the integration of the governing equations $m+1$ times, and does not require the computation of the Jacobian of the system.
We consider a nonlinear autonomous dynamical system in the form of 
\begin{equation}
\label{eq:dyn_syst}
    \mathbf{\dot{q}} = \mathbf{f(q)}
\end{equation}
where $\mathbf{q}$ is the system's state and $\mathbf{f}$ is a nonlinear operator. In chaotic solutions, the norm of a perturbation $\mathbf{y}_i$, such that $\mathbf{\hat{q}}_i = \mathbf{\overline{q}} + \mathbf{y}_i$ with $\mathbf{y}_i \ll 1$, grows in time until nonlinear saturation. For small enough times, $t_1 - t_0$, so that we avoid nonlinear saturation, the evolution of  $\mathbf{y}_i$ can be computed as
\begin{equation}
\label{eq: pert}
    \mathbf{y}_i (t_1) = \mathbf{\overline{q}}(t_1) - \mathbf{\hat{q}}_i(t_1),
\end{equation}
where both elements in the right-hand side are computed by solving \eqref{eq:dyn_syst} with initial conditions equal to $\mathbf{\overline{q}}(t_0)$ and $\mathbf{\overline{q}}(t_0)+\mathbf{y}_i(t_0)$, respectively. The average exponential growth rate for the perturbation $\mathbf{y}_i$ between $t_0$ and $t_1$ is
\begin{equation}
    \lambda = \frac{1}{t_1 - t_0}\ln\left(\frac{||\mathbf{y}(t_1)||}{||\mathbf{y}(t_0)||}\right),
\end{equation}
where $||\cdot||$ indicates the $L_2$ norm.
For $t_1 \to \infty$, almost all perturbations evolve with the same $\Lambda_1$, the dominant Lyapunov exponent 
\begin{equation}
    \Lambda_1 = \lim_{t_1\to\infty}\frac{1}{t_1-t_0}\ln\left(\frac{||\mathbf{y}(t_1)||}{||\mathbf{y}(t_0)||}\right),
\end{equation}
as the component along the direction with maximum growth becomes dominant for sufficiently long times. However, due to saturation of the nonlinear equations (or instability of the linearized equations), computing $\Lambda_1$ is not straightforward.

To compute the growth along the $m$ most unstable directions for long times, \citet{benettin1980lyapunov} proposed to periodically orthonormalize the evolution of the subspace spanned by $m$ different perturbations. The algorithm works as follows. Every $t_{\mathrm{o}}$, we orthonormalize the $m$ perturbations and compute the future evolution of the orthonormalized basis:
\begin{align}
    \mathbf{\Tilde{y}}_1(t) &= \frac{\mathbf{y}_1(t)}{||\mathbf{y}_1(t)||} \quad \mathrm{where} \quad \mathbf{y}_1(t-t_{\mathrm{o}}) = \epsilon \mathbf{\Tilde{y}}_1(t-t_{\mathrm{o}}), \nonumber \\ 
    \vdots &  \nonumber \\
    \mathbf{\Tilde{y}}_i (t) &= \frac{\mathbf{y'}_i(t)}{||\mathbf{y'}_i(t)||}; \quad \mathbf{y'}_i = \mathbf{y}_i - \sum_{j=1}^{i-1} (\mathbf{y}_i^T\mathbf{\Tilde{y}}_j) \mathbf{\Tilde{y}}_j \quad \mathrm{where} \quad \mathbf{y}_i(t-t_{\mathrm{o}}) = \epsilon\mathbf{\Tilde{y}}_i(t-t_{\mathrm{o}}), \nonumber \\
    \vdots &  \nonumber \\
    \mathbf{\Tilde{y}}_m (t) &= \cdots 
\end{align}
where $\epsilon \ll 1$ is selected in order for initial condition to be infinitesimal and $\mathbf{y}_i(t)$ is computed using \eqref{eq: pert}. For the first orthonormalization, the initial condition of the perturbations is random.
At each orthonormalization, we store the average exponential growths, so that for the $i$-th direction at the $k$-th orthonormalization we have 
\begin{equation}
    \lambda_i^{(k)} = \frac{1}{t_{\mathrm{o}}}\ln\left(\frac{||\mathbf{y}'_i(t)||}{||\mathbf{y}_i(t-t_{\mathrm{o}})||}\right)
\end{equation}
where $||\mathbf{y}_i(t-t_{\mathrm{o}})||=\epsilon$. After $N_\mathrm{o}$ orthonormalizations, the Lyapunov exponents are the average of the stored exponential growths
\begin{equation}
    \Lambda_i = \frac{1}{N_\mathrm{o}}\sum_{k=1}^{N_\mathrm{o}}\lambda_i^{(k)}.
\end{equation}


\subsection{Quasiperiodic dataset}
\label{app:qp_data}
In this section, we provide additional analysis on the saturation of the NRMSE of the autoencoder in the quasiperiodic case, panel (a) in Fig. \ref{fig:Rec}. 

\begin{figure}
    \centering
    \includegraphics[width=.66\textwidth]{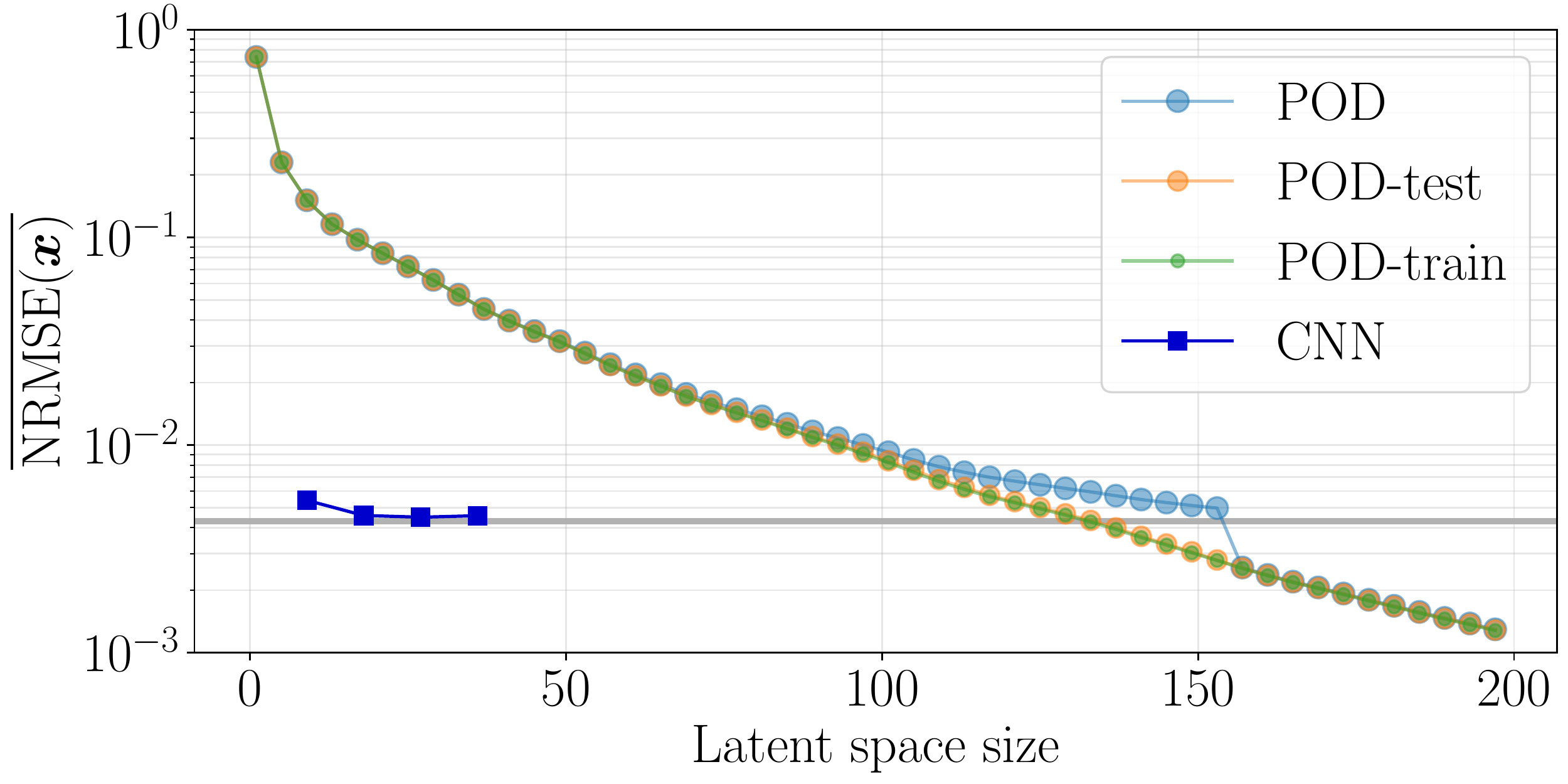}
    \caption{Discrepancy between training and test set in the quasiperiodic case.}
    \label{fig:QP_dataset}
\end{figure}

Figure \ref{fig:QP_dataset} shows the NRMSE for the autoencoder (CNN) and different POD projections, which differ one from the other for the training and test set used. The CNN and POD errors are obtained by generating the subspace using the training set and then computing the NRMSE in the test set. The POD-train and POD-test are obtained by generating the reduced order model on the training and test sets, respectively, and then computing the NRMSE on the same set that was used to generate the reduced order model. We observe that CNN and POD both shows a saturation of the NRMSE, while the POD-test and POD-train do not. This indicates that there is a discrepancy between training data and the test data. We observed the same saturation behavior for different training set of 30000 time units (results not shown), and we conclude that the saturation is due to the fact that a training set of 300000 time units is not fully representative of the entire trajectory of the system. This is a common situations in real world scenarios, where the available data is often incomplete.

\subsection{Extended results for the POD modes}

\begin{figure}
    \centering
    \includegraphics[width=1.\textwidth]{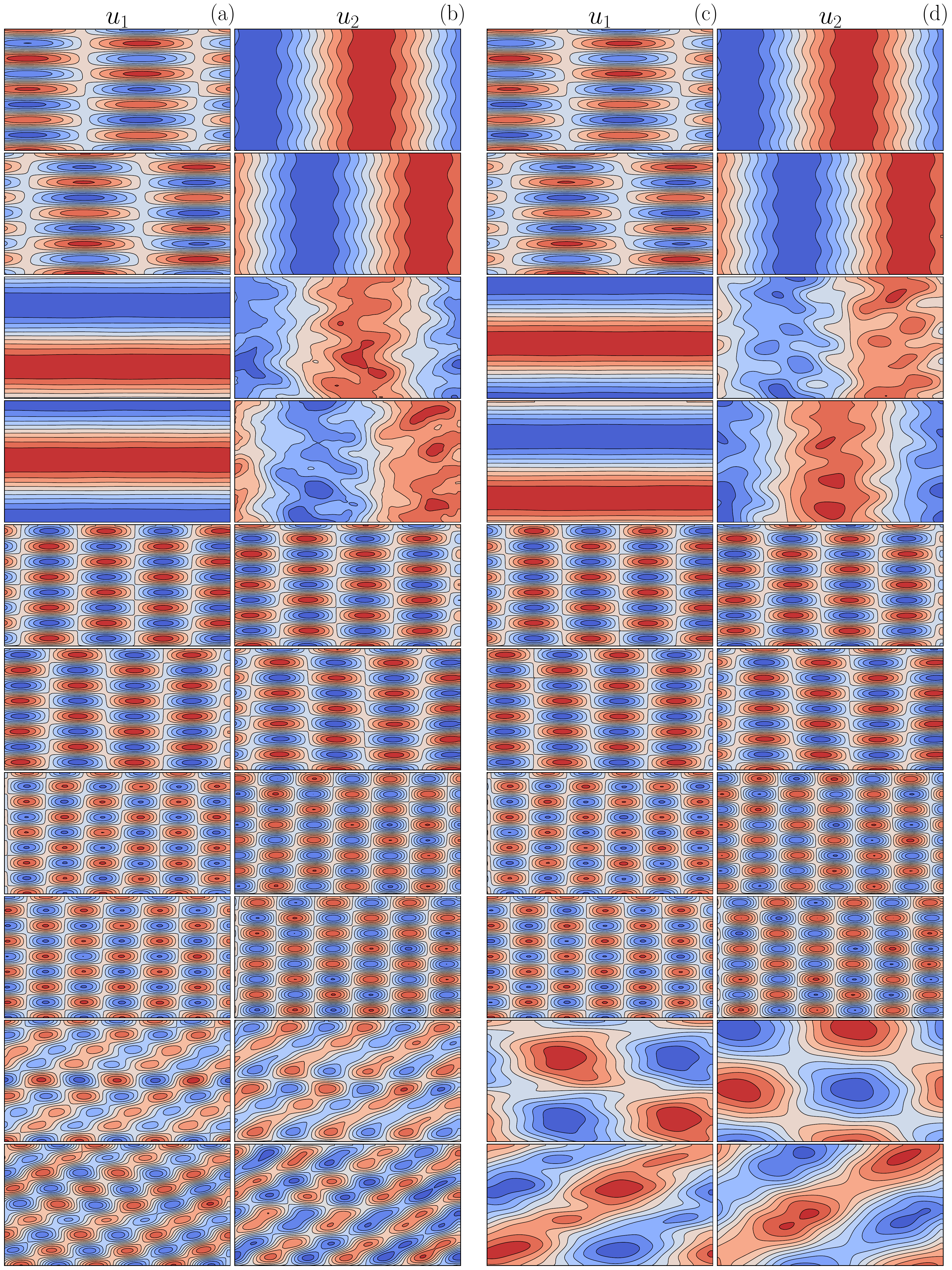}
    \caption{First ten POD modes, one per row, for the reconstructed flow field based on 4 POD modes in latent space  (a)-(b) and the true flow field (c)-(d). POD modes are sorted in descending order with first row being the first mode.}
    \label{fig:pods_10}
\end{figure}

In Figure \ref{fig:pods_10}, we show the comparison of Figure \ref{fig:pods} for the first ten POD modes.

\subsection{Dynamical content in the autoencoder loss function}
\label{app:der-train}
We provide an example of latent space tailored for the prediction in time of the system. 
We included in the autoencoder loss function the time-derivative of the system, which did not improve the results shown in \S\S \ref{sec:time-acc}-\ref{sec:statistics}. We report the formulation here to (i) prevent fellow researchers from trying what we observed not to work and to (ii) provide a basis for the further development of dynamically-aware reduced-order modelling with machine learning in fluids. 
We minimise the error on the time derivative of the system to include information about the dynamical content of the system during the training of the autoencoder:
\begin{equation}
    \mathcal{L} = \alpha_1 \sum^{N_t}_{i=1}\frac{1}{N_t}
    \frac{1}{ N_{\mathrm{phys}}}
    \left\vert\left\vert\hat{\boldsymbol{q}} - \boldsymbol{q}\right\vert\right\vert^2 + \alpha_2 \sum^{N_t}_{i=1}\frac{1}{N_t}
    \frac{1}{ N_{\mathrm{phys}}}
    \left\vert\left\vert\frac{\mathrm{d}\hat{\boldsymbol{q}}}{\mathrm{d}t} - \frac{\mathrm{d}\boldsymbol{q}}{\mathrm{d}t}\right\vert\right\vert^2,
\label{eq:der-loss}
\end{equation}
where $\alpha_1$ and $\alpha_2$ are normalization coefficients selected for the two losses to have the same order of magnitude. They are equal to  $1/\sigma( N_{\mathrm{phys}}^{-1} \left\vert\left\vert \hat{\boldsymbol{q}} - \boldsymbol{q} \right\vert\right\vert^2 )$ and $1/\sigma( N_{\mathrm{phys}}^{-1} \left\vert\left\vert \frac{d\hat{\boldsymbol{q}}}{dt} - \frac{d\boldsymbol{q}}{dt} \right\vert\right\vert^2)$, respectively; $\sigma(\cdot)$ is the standard deviation computed over the training set.
In this work, the time-derivative is computed using by first-order finite difference:
\begin{equation}
    \frac{\mathrm{d}\boldsymbol{q}}{\mathrm{d}t}(t_i) = \frac{\boldsymbol{q}(t_i + \delta t) - \boldsymbol{q}(t_i)}{\delta t},
\end{equation}
where $\delta t$ is defined in \S \ref{sec:data}. Other possible choices are the automatic differentiation \citep{racca2020automatic} or governing equations (if known). The purpose of minimising the error on the time-derivative is for the autoencoder to reconstruct not only the structures that account for the majority of the energy, but also the structures that contribute to the change in time of the state of the system.


\end{document}